\documentclass[aps,prd,reprint,preprintnumbers,showpacs]{revtex4-1}
\usepackage{graphicx}
\usepackage{amsmath}
\usepackage{caption}
\usepackage[section] {placeins}
\usepackage{slashed}
\usepackage{color}
\usepackage{soul}
\usepackage{appendix}
\usepackage{placeins}
\begin{document}

\title{ Why $\Xi(1690)$ and $\Xi(2120)$ are so narrow}
\author{K.~P.~Khemchandani$^{1,2}$\footnote{kanchan.khemchandani@unifesp.br}}
\author{A.~Mart\'inez~Torres$^{3}$\footnote{amartine@if.usp.br}}
\author{ A.~Hosaka$^4$\footnote{hosaka@rcnp.osaka-u.ac.jp}}
\author{ H.~Nagahiro$^{4,5}$\footnote{nagahiro@rcnp.osaka-u.ac.jp} }
\author{F.~S.~Navarra$^3$\footnote{navarra@if.usp.br}}
\author{ M.~Nielsen$^3$\footnote{mnielsen@if.usp.br}}
\preprint{}

 \affiliation{
$^1$ Departamento de Ci\^encias Exatas e da Terra, Universidade Federal de S\~ao Paulo, Campus Diadema, Rua Prof. Artur Riedel, 275, Jd. Eldorado, 09972-270, Diadema, SP, Brazil. \\
$^2$Faculdade de Tecnologia, Universidade do Estado do Rio de 
Janeiro, Rod. Presidente Dutra Km 298, P\'olo Industrial, 27537-000 , 
Resende, RJ, Brasil.\\
$^3$ Instituto de F\'isica, Universidade de S\~ao Paulo, C.P 66318, 05314-970 S\~ao Paulo, SP, Brazil.\\
$^4$ Research Center for Nuclear Physics (RCNP), Mihogaoka 10-1, Ibaraki 567-0047, Japan.\\
$^5$ Department of Physics, Nara Women's University,  Nara 630-8506, Japan.
}

\date{\today}

\begin{abstract}
The $\Xi$ baryons are expected to be naturally narrower as compared to their nonstrange and strange counterparts since they have only one light quark and, thus, their decay involves producing either a light meson and doubly strange baryon or both meson and baryon with strangeness which involves, relatively,  more energy. In fact, some $\Xi$'s  have full widths of the order of  even 10-20 MeV  when, in principle, they have a large phase space to decay to some open channels. Such is the case of $\Xi (1690)$, for which the width has been found to be of the order of 10 MeV in the latest {\it BABAR} and BELLE data.  In this manuscript we study why some $\Xi$'s are so narrow. Based on a coupled channel calculation of the pseudoscalar meson-baryon and vector meson-baryon systems with chiral and hidden local symmetry Lagrangians,  we find that the answer lies in the intricate  hadron dynamics. We find that the known mass, width, spin-parity and  branching ratios of  $\Xi (1690)$ can be naturally explained in terms of coupled channel meson-baryon dynamics.  We find another narrow resonance which can be related to  $\Xi(2120)$. We also look for exotic states $\Xi^{+}$ and $\Xi^{--}$ but find none. In addition we provide the cross sections for $ \bar K  \Lambda, \bar K \Sigma \rightarrow \pi \Xi$ which can be useful for understanding the enhanced yield of  $\Xi$ reported in recent studies of heavy ion collisions.
\end{abstract}

\maketitle

\section{Introduction}
Little is known about $\Xi$ resonances \cite{pdg} since it is difficult to produce them in the laboratory directly. There are limited facilities around the world which have anti-kaon beams and a  beam  of  nonstrange nature on a nucleon leads to low yields of  $\Xi$ baryons  as two pairs of strange quarks are required to be produced in this case. However, efforts are being made to improve this situation. For example, the photoproduction of $\Xi$'s on nucleons is being explored currently at the Jefferson laboratory \cite{clasx} considering the possibility of the production of a  hyperon resonance in the intermediate state, thus producing the strange quark pairs in two steps. Also, more information is expected to come in the future from the J-PARC \cite{jparc} and $\overline{\rm P}$ANDA \cite{panda} facilities.

In addition, studies made by BELLE  \cite{belle} and {\it BABAR} \cite{babar} collaborations show that  it is possible to extract useful information on this subject from rare processes  too. Specifically, some intriguing findings related to the properties of $\Xi(1690)$ have been reported in such studies. We find it useful to list these findings, and other relevant information on $\Xi(1690)$, in a separate subsection dedicated to $\Xi(1690)$ below. 
 
\subsection{$\Xi(1690)$}
The first observation of $\Xi(1690)$ in the $\bar K \Sigma$ invariant mass spectrum was reported in Ref.~\cite{Dionisi:1978tg}, where the mass and width were determined to be  $M = (1694 \pm 6)$~MeV, $\Gamma = (26 \pm 6)$~MeV in the negatively charged channel and $M = (1684 \pm 5)$~MeV, $\Gamma = (20 \pm 4)$~MeV in the neutral channel.  The spin-parity of this state was not determined in Ref.~\cite{Dionisi:1978tg}. Although the mass values obtained by the latest investigations \cite{belle,babar} are not very different, the width of $\Xi(1690)$ has been determined to be even narrower, of the order of 10 MeV \cite{belle,babar}. To mention explicit results, the {\it BABAR} collaboration deduced the mass and width of $\Xi(1690)$ to be: $M~=~\left(1684.7 \pm 1.3^{+2.2}_{-1.6} \right)$~MeV, $\Gamma = \left(8.1^{+3.9+1.0}_{-3.5-0.9}\right)$ MeV \cite{babar}, and the BELLE Collaboration reported: $M  = \left(1688 \pm 2\right)$ MeV, $\Gamma = \left(11 \pm 4\right)$ MeV \cite{belle}. Both studies obtained spin-parity of this state as $1/2^-$. It was also found in Ref.~\cite{babar} that a clear evidence for $\Xi^0(1690)$ is found in the $\bar K^0 \Lambda$ invariant mass spectrum but not in the $\pi^+ \Xi^-$ spectrum.  The information on the 
 ratio of $\Xi^0(1690)$ to $\bar K \Lambda$ and $\bar K \Sigma$ has been updated in Ref.~\cite{belle} to, 
\begin{equation}
\frac{B(\Xi^0(1690) \to K^- \Sigma^+)}{B(\Xi^0(1690) \to \bar K^0 \Lambda)} = 0.50 \pm 0.26. \label{ratio}
\end{equation}
as compared to the value ($1.8 \pm 0.6$) obtained from the older analysis of Ref.~\cite{Dionisi:1978tg}.

\begin{figure}[h!]
\includegraphics[width=0.3\textwidth]{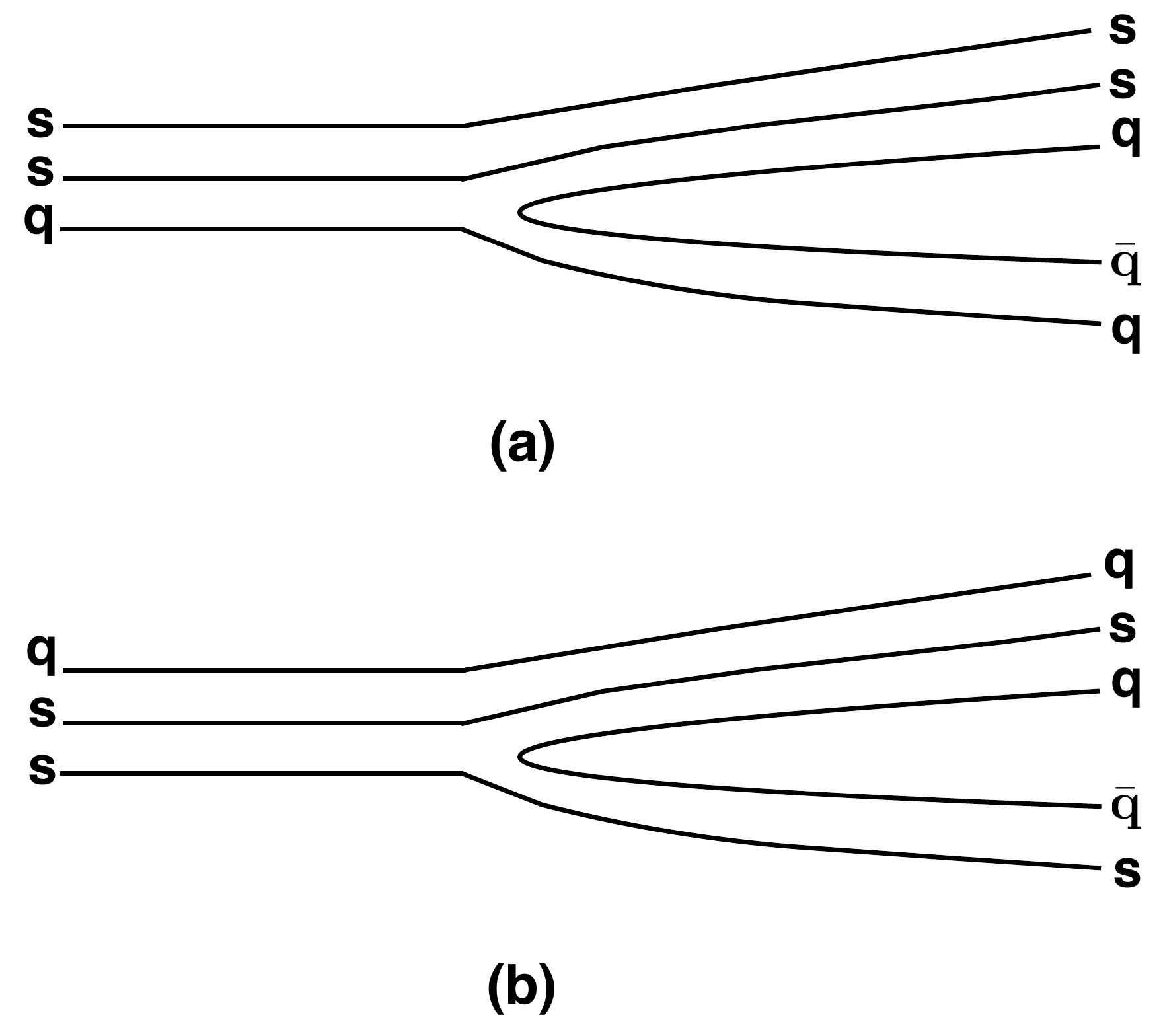}
\caption{Decay of a $\Xi$ resonance to a meson ($q\bar q$/$s \bar q$)-baryon ($qss$/$qqs$) channel.}\label{decay}
\end{figure}

These properties are difficult to understand, for example, the narrow width of $\Xi(1690)$ in spite of having a large phase space to decay to open channels, like, $\pi \Xi$ and  $\bar K\Lambda$.  Similar is the case of the finding related to the suppressed decay to $\pi \Xi$,  as compared to the decay to the $\bar K \Lambda$ channel (as is evident from the difficulty in the identification of the signal of $\Xi^0(1690)$ in the  $\pi^+ \Xi^-$ spectrum of Refs.~\cite{babar,belle}). A light quark pair production, within a na\"ive quark model, is equivalently required in both cases (as can be seen in Fig.~\ref{decay}).
%It is also interesting that  $\Xi(1690)$, being a spin 1/2 particle, has a higher mass than the first excited state of the spectrum $\Xi(1530)$ which has a spin $3/2$ \cite{pdg}. There is a case of ``parity-inverted" states ($N^*(1440)$ and $N^*(1535)$) in the nucleon spectrum, which can be explained in terms of a five quark content \cite{zou_npa,juan,bruns,more}, but $\Xi(1530)$ and $\Xi(1690)$ seem to have an inverted spin.  
The branching fraction to the $\bar K\Sigma$ and $\bar K\Lambda$ channels (shown in Eq.~(\ref{ratio}))  is also counterintuitive: there is little phase space for decay to the former one and one would anticipate this ratio to be much smaller. 

In fact, these experimental findings have motivated theoretical investigations within quark models, and the mass has been obtained: $M \sim$ 1725 MeV \cite{Pervin:2007wa}.  Also, the decay properties of $\Xi(1690)$ have been studied within a chiral quark model \cite{Xiao:2013xi}, which includes hadron dynamics through a quark-meson effective coupling, and its total width was found to be 25~MeV. This value, although not widely different, is larger than the one observed in Refs.~\cite{babar, belle}. Recently, the findings of the BES and the BELLE Collaborations motivated a study of $\Xi(1690)$  in terms of hadron degrees of freedom also \cite{Sekihara}. The formalism used in Ref.~\cite{Sekihara}, based on a $t$-channel interaction of pseudoscalar meson-baryon channels, reproduces well the mass of $\Xi(1690)$ but the width is found to be around 1 MeV. The author of Ref.~\cite{Sekihara} suggested that inclusion of more diagrams might be useful in obtaining a width of the order of 10 MeV.  

In view of the growing interest in the $\Xi$-baryons at the experimental facilities, %and the recent puzzling findings, 
we find it timely to study cascade resonances in meson-baryon coupled systems. Our formalism includes interaction of both pseudoscalar as well as vector mesons, and in the latter case we include $s$-,~$t$-,~$u$-channel diagrams and a contact term, whose origin will be explained in a subsequent section. As we shall show in this article, in agreement with Ref.~\cite{Sekihara}, $\Xi(1690)$ can be understood as a state generated due to meson-baryon coupled channel dynamics and  it is possible to straightforwardly  understand all the experimental results reported in Refs.~\cite{babar,belle}.  
 A hadronic dynamical origin of resonances was predicted long time ago by T\"ornqvist~\cite{Tornqvist:1991ks} and the possibility of understanding some $\Xi$ baryons (including $\Xi(1690)$) in terms of hadron dynamics has been discussed earlier in the literature ~\cite{ramosXi,Kolomeitsev:2003kt,GarciaRecio:2003ks,ramosvb,Gamermann, Pavon,Nakayama,javi,Oh}. We will compare our results  with those obtained in such previous studies in this manuscript.

As we shall discuss, we find an evidence for another $\Xi$ resonance, which is also narrower than expected (due to the existence of several open channels for decay) and can be related to $\Xi(2120)$.  We find that the nature of this state is such that it can be difficult to identify its presence in the experimental data related to the $\bar K \Lambda$ mass spectrum.  We suggest better channels to investigate the properties of $\Xi(2120)$.

\subsection{$\Xi(2120)$}
Before proceeding further, for the sake of completeness, we find it useful to present the known information on $\Xi(2120)$ as well. Not much information is available on this resonance. Its spin-parity is not known. This state was first found  in $\bar K \Lambda$ mass spectrum in Ref.~\cite{Gay:1976sr}, which reported its mass and width values to be $M = 2123 \pm 7$ MeV and $\Gamma = 25 \pm 12$ MeV. One of the later investigation confirmed an accumulation of events around 2120 MeV in $\bar K \Lambda$ and determined the mass and width to be $M = 2137 \pm 4$ MeV and $\Gamma \sim 20$ MeV \cite{Vorobev:1979as}. This state was, however, not found in Ref.~\cite{Hemingway:1977uw}.
All these experimental data suffer from poor statistics. A better quality data is required to determine the properties of $\Xi$'s with mass above 2 GeV.  Indications of the existence of  $\Xi$ resonances around 2120 MeV, whose origin  lies in the meson-baryon coupled channel dynamics, has been discussed in some previous works. For example, in Ref.~\cite{Gamermann}, a pole around 2100 MeV is found with a full width of $\sim 100$ MeV for a $\Xi$, with $J^\pi = 1/2^-$ and $ 3/2^-$, as a result of pseudoscalar and vector meson-baryon interaction treated within a $SU(6)$ symmetry. In Ref.~\cite{ramosvb} also a $\Xi$ with $J^\pi = 1/2^-$ and $ 3/2^-$, with mass $\sim$ 2100 MeV and width $\sim 60$ MeV is found to get generated from meson-baryon dynamics. As we will show, we find the width of $\Xi(2120)$ to be narrower.

We discuss the details of our formalism and results in the next sections. In the discussions on the results, we will show that the presence of  resonances affects  the cross sections of processes like $\bar K \Lambda \to \pi \Xi$ and  $\bar K \Sigma \to \pi \Xi$, which can be important in understanding the enhanced $\Xi$ production found in the Ar + KCl collisions by the HADES collaboration \cite{hades}.

\section{Formalism}

The formalism of the present study is based on solving the Bethe-Salpeter equation in a coupled channel approach. To do this we consider all systems with strangeness $-2$ formed by a pseudoscalar/vector meson with an octet baryon: $\pi \Xi$,  $\eta \Xi$, $\bar K \Sigma$, $\bar K \Lambda$, $\rho \Xi$,  $\omega \Xi$, $\phi \Xi$, $\bar K^* \Sigma$ and $\bar K^* \Lambda$. 

We obtain the amplitudes for different transitions among these channels using Lagrangians based on effective field theories. To determine the vector-meson--baryon interaction we use the formalism developed in our previous work \cite{vbvb}. A detailed analysis of the low energy vector meson interaction with octet baryons was carried out in Ref~\cite{vbvb} by calculating $s$-, $t$- and $u$-channel diagrams and a contact interaction, using a Lagrangian invariant under the gauge transformations of the hidden-local symmetry. It was found that the contribution of all the diagrams is of comparable size and that the full (summed) amplitude depended on the total spin as well as the isospin of the system. Following Ref.~\cite{vbvb}, thus, we write vector-baryon (VB) amplitudes for each spin-isospin  configuration as
\begin{align}
V^{I,S}_{\textrm VB} = V^{I,S}_{t, {\textrm VB}} + V^{I,S}_{s,{\textrm VB}} + V^{I,S}_{u,{\textrm VB}} + V^{I,S}_{\textrm{CT,VB}}.\label{VB}
\end{align}
These amplitudes can be obtained from the general Lagrangian
\begin{eqnarray} \label{vbb}
&&\mathcal{L}_{\textrm VB}= -g \Biggl\{ \langle \bar{B} \gamma_\mu \left[ V_8^\mu, B \right] \rangle + \langle \bar{B} \gamma_\mu B \rangle  \langle  V_8^\mu \rangle  
\Biggr. \\ \nonumber
&&+ \frac{1}{4 M} \left( F \langle \bar{B} \sigma_{\mu\nu} \left[ V_8^{\mu\nu}, B \right] \rangle  + D \langle \bar{B} \sigma_{\mu\nu} \left\{ V_8^{\mu\nu}, B \right\} \rangle\right)\\\nonumber
&& +  \Biggl.  \langle \bar{B} \gamma_\mu B \rangle  \langle  V_0^\mu \rangle  
+ \frac{ C_0}{4 M}  \langle \bar{B} \sigma_{\mu\nu}  V_0^{\mu\nu} B  \rangle  \Biggr\},
\end{eqnarray}
where  $\langle ... \rangle$ refers to an $SU(3)$ trace, the subscript $8$ ($0$) on the meson fields denotes the octet (singlet) part of their wave function (relevant in case of $\omega$ and $\phi$, for which we assume an ideal mixing). $V^{\mu\nu}$ represents the  tensor field of the vector mesons,
\begin{equation}
V^{\mu\nu} = \partial^{\mu} V^\nu - \partial^{\nu} V^\mu + ig \left[V^\mu, V^\nu \right], \label{tensor}
\end{equation}
 and $V^\mu$ and $B$ denote the SU(3) matrices for the (physical) vector mesons and octet baryons
\begin{eqnarray}
V^\mu &=&\frac{1}{2}
\left( \begin{array}{ccc}
\rho^0 + \omega & \sqrt{2}\rho^+ & \sqrt{2}K^{*^+}\\
&& \\
\sqrt{2}\rho^-& -\rho^0 + \omega & \sqrt{2}K^{*^0}\\
&&\\
\sqrt{2}K^{*^-} &\sqrt{2}\bar{K}^{*^0} & \sqrt{2} \phi 
\end{array}\right)^\mu
, \\
B &=&
\left( \begin{array}{ccc}
 \frac{1}{\sqrt{6}} \Lambda + \frac{1}{\sqrt{2}} \Sigma^0& \Sigma^+ & p\\
&& \\
\Sigma^-&\frac{1}{\sqrt{6}} \Lambda- \frac{1}{\sqrt{2}} \Sigma^0 &n\\
&&\\
\Xi^- &\Xi^0 & -\sqrt{\frac{2}{3}} \Lambda 
\end{array}\right).
\end{eqnarray}
In Eq.(\ref{vbb}), the coupling $g$ is related to meson decay constants as
\begin{align}
g=\frac{m}{\sqrt{2}f}, \label{ksrf}
\end{align}
and the constants $D$ = 2.4, $F$ = 0.82 and $C_0 = 3F - D$ are such that  the anomalous magnetic 
couplings of $\rho NN$, $\omega NN$ and $\phi NN$ vertices are correctly reproduced. These values have also been found useful in calculations of the magnetic moments of the baryons in Ref.~\cite{jidohosaka}. 

 Keeping in mind that the thresholds of different VB channels differ by $\sim$ 200 MeV, which implies that the center of mass energies of the lighter VB channels can vary up to 200 MeV above the respective thresholds, we calculate all the amplitudes relativistically, following Refs.~\cite{Penner:2002ma,Penner:2002md}. 
 
 We start the discussion on different amplitudes with the  contact interaction which arises from the commutator part of the tensor field (Eq.~(\ref{tensor})). The resulting amplitude has a form,
\begin{eqnarray}\label{contact}
&&V^{I}_{\textrm{CT,VB}}  = - C_{\rm CT, VB}^{I} \frac{g_1 g_2}{2 \sqrt{M_1 M_2}} \left\{ - i \vec{\sigma}\cdot \vec{\epsilon}_2 \times \vec{\epsilon}_1  \right. \\\nonumber
&+& \frac{1}{E_2 + M_2} \left(-\epsilon_1^0 \vec{\sigma}\cdot\vec{P}_2  \vec{\sigma}\cdot\vec{\epsilon}_2  + \epsilon_2^0 \vec{\sigma}\cdot\vec{P}_2  \vec{\sigma}\cdot\vec{\epsilon}_1 \right) \\\nonumber
&+&\frac{1}{E_1 + M_1} \left(-\epsilon_2^0  \vec{\sigma}\cdot\vec{\epsilon}_1 \vec{\sigma}\cdot\vec{P}_1  + \epsilon_1^0 \vec{\sigma}\cdot\vec{\epsilon}_2 \vec{\sigma}\cdot\vec{P}_1   \right) \\\nonumber
&-& \frac{1}{2\left(E_1 + M_1\right)\left(E_2 + M_2\right)}\left( \vec{\sigma}\cdot\vec{P}_2\vec{\sigma}\cdot\vec{\epsilon}_1\vec{\sigma}\cdot\vec{\epsilon}_2 \vec{\sigma}\cdot\vec{P}_1\right.\\\nonumber
&-&\left. \left.  \vec{\sigma}\cdot\vec{P}_2\vec{\sigma}\cdot\vec{\epsilon}_2\vec{\sigma}\cdot\vec{\epsilon}_1 \vec{\sigma}\cdot\vec{P}_1\right)\right\} \sqrt{\frac{M_1 + E_1}{2 M_1}} \sqrt{\frac{M_2 + E_2}{2 M_2}},
\end{eqnarray}
%\end{widetext}
where $M_1$, ($M_2$), $E_1$ ($E_2$), $\vec{P}_1$ ($\vec{P}_2$) represent the mass, energy and the three-momentum of the baryon in the initial (final) state, respectively, and $g_1$ ($g_2$) is related to the decay constant of the vector meson in the initial (final) state  through Eq.~(\ref{ksrf}).  The  values of the decay constants used in our work are: $f_\pi$ = 93 MeV, $f_\eta$ = 120.9 MeV, $f_k$ = 113.46 MeV, $f_\rho$, $f_\omega$ = 153.45, $f_\phi$ = 168.33 MeV, $f_{K^*}$ = 159.96 MeV \cite{Maris:1999nt,Maris:1999sz}. In Eq.~(\ref{contact}), %$\epsilon_1 = \left(\epsilon_1^0, \vec{\epsilon}_1 \right)$, $\epsilon_2 = \left(\epsilon_2^0, \vec{\epsilon}_2 \right)$ 
$\epsilon_1^0$ ($\epsilon_2^0$) and $\vec{\epsilon}_1$ ($\vec{\epsilon}_2$)
represent the temporal and the spatial part of the polarization four-vectors of the mesons in the initial (final) state, respectively, and $C^{I}_{\textrm{CT,VB}}$ are isospin dependent coefficients whose values are given in the Appendix~\ref{coeff}, in Tables~\ref{vbvbCT1/2},~\ref{vbvbCT3/2} for the different relevant channels. We recall that the main purpose of the present article is to study the formation of resonances in a coupled meson-baryon system. We are, thus, interested in low energy dynamics of such a system and, consider the s-wave contribution of different amplitudes. %and, as is customary, obtain all the amplitudes within a nonrelativistic approach. 
The s-wave part ($l = 0$) of Eq.~(\ref{contact}) gives 
{\small
\begin{eqnarray}\label{contacts}
&&V^{I,S}_{\textrm{CT,VB}} = - C_{\rm CT, VB}^{I} \frac{g_1 g_2}{2 \sqrt{M_1 M_2}}\sqrt{\frac{M_1 + E_1}{2 M_1}} \sqrt{\frac{M_2 + E_2}{2 M_2}} \\\nonumber
&& \times \biggl\{ - i \vec{\sigma}\cdot \vec{\epsilon}_2 \times \vec{\epsilon}_1 +  \frac{1}{E_1 + M_1} \left(-\epsilon_2^0  \vec{\sigma}\cdot\vec{\epsilon}_1 \vec{\sigma}\cdot\vec{P}_1  + \epsilon_1^0 \vec{\sigma}\cdot\vec{\epsilon}_2 \vec{\sigma}\cdot\vec{P}_1   \right) \biggr\},
\end{eqnarray}
}
which is indicated by the superscript $S$ representing the spin  of the VB system (which coincides with the total angular momentum $J$ of the sysem).

The amplitude of Eq.~(\ref{contacts}) is projected on the  total spin half base, to get,
\begin{eqnarray}\nonumber
&&V^{I,1/2}_{\textrm{CT,VB}} = C_{\rm CT, VB}^{I} \frac{g_1 g_2}{ \sqrt{M_1 M_2}}\sqrt{\frac{M_1 + E_1}{2 M_1}} \sqrt{\frac{M_2 + E_2}{2 M_2}}  \\
&&\times\left\{ 1 +  \left(\frac{1}{E_1 + M_1} \right)\left(\frac{\mid \vec{K}_1\mid^2}{2 m_1} - \frac{\mid \vec{K}_1\mid\mid \vec{K}_2\mid}{6 m_2}\right)\right\},\label{CT1h}
\end{eqnarray}
and in case of spin 3/2, we obtain
\begin{eqnarray}\nonumber
&&V^{I,3/2}_{\textrm{CT,VB}}  = -C_{\rm CT, VB}^{I}\frac{g_1 g_2}{ 2 \sqrt{M_1 M_2}} \sqrt{\frac{M_1 + E_1}{2 M_1}} \sqrt{\frac{M_2 + E_2}{2 M_2}} \\
&&\times \biggl\{ 1 +  \left(\frac{1}{E_1 + M_1}\right) \left(-\frac{4 \mid \vec{K}_1\mid \mid \vec{K}_2\mid}{3 m_2}\right)\biggr\}.\label{CT3h}
\end{eqnarray}
In Eqs.~(\ref{CT1h}) and (\ref{CT3h}), (and throughout this manuscript) $m_1$, $m_2$ and $\vec{K}_1$, $\vec{K}_2$ represent the masses and the three-momenta of the vector mesons in the initial and final state, respectively.\\

Using the Yukawa-type vertices  obtained from Eq.~(\ref{vbb}), we deduce the $s$- and $u$-channel amplitudes by treating the vector mesons relativistically, and obtain, 
{\small
\begin{widetext}
\begin{eqnarray}\nonumber
V^{I}_{s, {\textrm VB}} &=& \frac{g_1 g_2}{s - M_x^2}\left\{\left(\sqrt{s} + M_x\right)\left(\epsilon^0_2\left[I^s_{1f} + \frac{I^s_{2f}K_2^0}{2\sqrt{M_1 M_2}}\right] -\frac{I^s_{2f} \left(\vec{\sigma}\cdot\vec{\epsilon}_2\vec{\sigma}\cdot\vec{K}_2\right)}{2\sqrt{M_1 M_2}}\right)\left(\epsilon_1^0\left[I^s_{1i} +\frac{I^s_{2i} K_1^0}{2\sqrt{M_1 M_2}}\right]-\frac{I^s_{2i} \left( \vec{\sigma}\cdot\vec{K}_1\vec{\sigma}\cdot\vec{\epsilon}_1\right)}{2\sqrt{M_1 M_2}}\right)\right.\\
&&\left.+ \left(\sqrt{s} - M_x\right)\left(\vec{\sigma}\cdot\vec{\epsilon}_2\left[I^s_{1f}-\frac{I^s_{2f} K_2^0}{2\sqrt{M_1 M_2}}\right] +\frac{I^s_{2f}\epsilon_2^0 \left(\vec{\sigma}\cdot\vec{K}_2 \right) }{2\sqrt{M_1 M_2}}\right)\left(\left[I^s_{1i} -\frac{I^s_{2i} K_1^0}{2\sqrt{M_1 M_2}}\right]\vec{\sigma}\cdot\vec{\epsilon}_1+\frac{I^s_{2i}\epsilon_1^0 \left(\vec{\sigma}\cdot\vec{K}_1\right)}{2\sqrt{M_1 M_2}}\right)\right\},\label{schannel}
\end{eqnarray}
\begin{eqnarray}\nonumber
V^{I}_{u, {\textrm VB}} &=& \frac{g_1 g_2}{u-M_x^2} \Biggl\{ \epsilon_1^0\epsilon_2^0 \left(\frac{1}{2}\left[E_2 -  K_1^0 +E_1 - K_2^0\right] + M_x\right)V_{u1} +\vec{\sigma}\cdot\vec{\epsilon}_1\vec{\sigma}\cdot\vec{\epsilon}_2 \left(\frac{1}{2}\left[ E_2 -  K_1^0 +E_1 - K_2^0\right] - M_x\right)V_{u2}  \Biggr.\\\nonumber &+& \epsilon_2^0~ \vec{\sigma}\cdot\vec{\epsilon}_1\vec{\sigma}\cdot\vec{K}_2~V_{u3} 
+ \epsilon_2^0~ \vec{\sigma}\cdot\vec{\epsilon}_1\vec{\sigma}\cdot\vec{K}_1~V_{u4}
+ \epsilon_1^0~ \vec{\sigma}\cdot\vec{K}_2\vec{\sigma}\cdot\vec{\epsilon}_2~ V_{u5}+\epsilon_1^0~ \vec{\sigma}\cdot\vec{K}_1\vec{\sigma}\cdot\vec{\epsilon}_2~ V_{u6} + \epsilon_1^0\epsilon_2^0 ~\frac{\vec{\sigma}\cdot\vec{K}_2\vec{\sigma}\cdot\vec{K}_2}{4\sqrt{M_1 M_2}}~V_{u7} \\\nonumber
&+&  \epsilon_1^0\epsilon_2^0 ~ \frac{\vec{\sigma}\cdot\vec{K}_1\vec{\sigma}\cdot\vec{K}_1}{4\sqrt{M_1 M_2}}~V_{u8}   + \epsilon_1^0\epsilon_2^0 ~ \frac{\vec{\sigma}\cdot\vec{K}_1\vec{\sigma}\cdot\vec{K}_2}{2 \sqrt{M_1 M_2}}~V_{u9}
+ \frac{\vec{\sigma}\cdot\vec{\epsilon}_1\vec{\sigma}\cdot\vec{K}_2\vec{\sigma}\cdot\vec{K}_2\vec{\sigma}\cdot\vec{\epsilon}_2}{4 \sqrt{M_1 M_2}}~V_{u10} + \frac{\vec{\sigma}\cdot\vec{\epsilon}_1\vec{\sigma}\cdot\vec{K}_1\vec{\sigma}\cdot\vec{K}_1\vec{\sigma}\cdot\vec{\epsilon}_2}{4 \sqrt{M_1 M_2}}~V_{u11}\\\nonumber
&+& \frac{\vec{\sigma}\cdot\vec{\epsilon}_1\vec{\sigma}\cdot\vec{K}_1\vec{\sigma}\cdot\vec{K}_2\vec{\sigma}\cdot\vec{\epsilon}_2}{2 \sqrt{M_1 M_2}}~V_{u12}-\frac{I^s_{2f}I^s_{2i}}{8 M_1 M_2} \Biggl[\epsilon_1^0~\vec{\sigma}\cdot\vec{K}_1\vec{\sigma}\cdot\vec{K}_1\vec{\sigma}\cdot\vec{K}_2\vec{\sigma}\cdot\vec{\epsilon}_2  + \epsilon_1^0 \vec{\sigma}\cdot\vec{K}_1\vec{\sigma}\cdot\vec{K}_2\vec{\sigma}\cdot\vec{K}_2\vec{\sigma}\cdot\vec{\epsilon}_2 \Biggr. \\
&+& \Biggl.\Biggl.\epsilon_2^0~\vec{\sigma}\cdot\vec{\epsilon}_1\vec{\sigma}\cdot\vec{K}_1\vec{\sigma}\cdot\vec{K}_1\vec{\sigma}\cdot\vec{K}_2 + \epsilon_2^0\vec{\sigma}\cdot\vec{\epsilon}_1\vec{\sigma}\cdot\vec{K}_1\vec{\sigma}\cdot\vec{K}_2\vec{\sigma}\cdot\vec{K}_2\Biggr]\Biggr\},\label{u-channel}
\end{eqnarray}
\end{widetext}}\noindent
where  $M_x$ is the mass of the exchanged (octet) baryon, $s$ and $u$ are the Mandelstam variables, $K_1^0$ ($K_2^0$) is the energy of the incoming (outgoing) vector meson, $I^s_{1i}$, $I^s_{1f}$, $I^s_{2i}$, $I^s_{2f}$ are the isospin coefficients for different channels whose products are given in the Appendix~\ref{coeff},  in Tables~\ref{i1fi1iS},~\ref{i1fi2iS},~\ref{i2fi1iS},~\ref{i2fi2iS}.  The definitions of $V_{u1}$, $V_{u2}$~$...$~$V_{u12}$ appearing in Eq.~(\ref{u-channel}) are also given in Eqs.~(\ref{u1}-\ref{u12}), in Appendix~\ref{conv}.  Due to the kinematic dependence of the Mandelstam variable $u$ on the incoming and outgoing momentum, the s-wave projection of the  $u$-channel amplitude is done numerically. In case of the $s$-channel amplitude we can project Eq.~(\ref{schannel}) analytically on s-wave, which gives
%\begin{eqnarray}
%%%V^I_u&=-C^I_u\left(\frac{g^2}{2\bar M-m}\right) ,\nonumber\\
%V^{I,S}_{s, {\textrm VB}}&=&C^I_{s, {\textrm VB}} \left(\frac{g_1 g_2}{2\bar M + \bar m}\right) \vec{\epsilon}_2\cdot \vec{\sigma}\,\, \vec{\epsilon}_1\cdot \vec{\sigma },\label{Vs}\\
%V^I_\textrm{CT}&=C^I_{\textrm{CT}}\,\frac{g^2}{M},\label{Vuc1}
%V^{I,S}_{u, {\textrm VB}}&=& C^I_{u, {\textrm VB}} \left(\frac{g_1 g_2}{2\bar M - \bar m}\right) \vec{\epsilon}_1\cdot \vec{\sigma}\,\, \vec{\epsilon}_2\cdot \vec{\sigma },\label{Vu}
%\end{eqnarray}
{\small
\begin{widetext}
\begin{eqnarray}\nonumber
&&V^{I,S}_{s, {\textrm VB}} = \frac{g_1 g_2}{s - M_x^2}\left\{\left(\sqrt{s} + M_x\right)\left(\epsilon^0_2\epsilon^0_1\left(I^s_{1f} + \frac{I^s_{2f}K_2^0}{2\sqrt{M_1 M_2}}\right)\left(I^s_{1i} + \frac{I^s_{2i}K_1^0}{2\sqrt{M_1 M_2}}\right) -\frac{\epsilon_2^0 I^s_{2i} \left( \vec{\sigma}\cdot\vec{K}_1\vec{\sigma}\cdot\vec{\epsilon}_1\right)}{2\sqrt{M_1 M_2}} \left(I^s_{1f} -\frac{I^s_{2f} K_2^0}{2\sqrt{M_1 M_2}}\right)\right)\right.\\
&&+\left.\left(\sqrt{s} - M_x\right)\left(\vec{\sigma}\cdot\vec{\epsilon}_2\vec{\sigma}\cdot\vec{\epsilon}_1\left(I^s_{1f}-\frac{I^s_{2f} K_2^0}{2\sqrt{M_1 M_2}}\right) \left(I^s_{1i}+\frac{I^s_{2i} K_1^0}{2\sqrt{M_1 M_2}}\right)+\frac{I^s_{2i}\epsilon_1^0 \left(\vec{\sigma}\cdot\vec{\epsilon}_2\vec{\sigma}\cdot\vec{K}_1\right)}{2M_1 M_2}\left(I^s_{1f} -\frac{I^s_{2f} K_2^0}{2M_1 M_2}\right)\right)\right\},\label{schannel2}
\end{eqnarray}
\end{widetext}}
which, in turn, on spin-half projection reduces to 
{\small\begin{widetext}
\begin{eqnarray}\nonumber
&&V^{I,1/2}_{s, {\textrm VB}} = \frac{g_1 g_2}{s - M_x^2}\left\{\left(\sqrt{s} + M_x\right)\left(\frac{|\vec{K}_1||\vec{K}_2|}{3 m_1 m_2}\left(I^s_{1f} + \frac{I^s_{2f}K_2^0}{2\sqrt{M_1 M_2}}\right)\left(I^s_{1i} + \frac{I^s_{2i}K_1^0}{2\sqrt{M_1 M_2}}\right) \right.
\right.\\\nonumber 
&-& \left. \frac{|\vec{K}_2||\vec{K}_1|}{m_2}\left(\frac{K_2^0}{3 m_2} +\frac{2}{3} \right)\frac{I^s_{2i}}{2\sqrt{M_1 M_2}} \left(I^s_{1f} -\frac{I^s_{2f} K_2^0}{2\sqrt{M_1 M_2}}\right)\right)\\
&&+\left.\left(\sqrt{s} - M_x\right)\left(3\left(I^s_{1f}-\frac{I^s_{2f} K_2^0}{2\sqrt{M_1 M_2}}\right) \left(I^s_{1i}-\frac{I^s_{2i} K_1^0}{2\sqrt{M_1 M_2}}\right)+\frac{|\vec{K}_1|^2}{m_1}\left(\frac{K_2^0}{3 m_2} +\frac{2}{3} \right)\frac{I^s_{2i}}{2\sqrt{M_1 M_2}}\left(I^s_{1f} -\frac{I^s_{2f} K_2^0}{2\sqrt{M_2 M_2}}\right)\right)\right\}.\label{schannel3}
\end{eqnarray}
\end{widetext}}
For more details on the conventions/normalizations related to the polarization vectors used in our work, we refer the reader to Appendix~\ref{conv}.
%SU(3) averaged masses of baryons and vector mesons, respectively, are used for $\bar M$ and $\bar m$, for practical purposes. To obtain these amplitudes we consider the exchange of octet baryons. Thus,  s-channel diagram (Eq.~(\ref{Vs})) is null  for the configurations with isospin 3/2 or spin 3/2. The values of the isospin coefficients $C^{1/2}_{s, {\textrm VB}}$ and $C^I_{u, {\textrm VB}}$ are given in the Appendix, in Table~\ref{vbvbs}, and Tables~\ref{vbvbu1/2}, \ref{vbvbu3/2}, respectively.

Finally, the contribution of  the $t$-channel amplitude is obtained as,
\begin{widetext}
\begin{eqnarray}\label{Vtex}
%V^{I}_{t, {\rm VB}} &=&  \frac{C_{t, {\rm VB}}^{I}}{4 f_V^2} \left(2 \sqrt{s} - M_1 - M_2 \right) \\
%&\times&\sqrt{\frac{M_1 + E_1}{2 M_1}} \sqrt{\frac{M_2 + E_2}{2 M_2}}\,\,\epsilon_1\cdot \epsilon_2.%-\frac{C^I_{t, VB}}{4 f_{V}^2}(\omega + \omega^\prime)  \vec{\epsilon}_1\cdot \vec{\epsilon}_2,
V^{I}_{t, {\rm VB}} &=& \frac{-m_{Vx}^2}{4 f_{Vi} f_{Vj}}\frac{1}{t-m_{Vx}^2}\left\{ \epsilon_1\cdot\epsilon_2 \left[ \left( 2 \sqrt{s} - M_1 - M_2 +  (M_1-M_2)\frac{(m_2^2-m_1^2)}{m_{Vx}^2} \right) C_{t1, {\rm VB}}^{I} \right. \right. \\\nonumber
&+& \left( \frac{M_1+ M_2}{2 M}\left(2 \sqrt{s} - M_1 - M_2\right)-\frac{s-u}{2M}\right)C_{t2, {\rm VB}}^{I}
+ \frac{\vec \sigma\cdot\vec P_2 ~\vec \sigma\cdot\vec P_1}{\left(E_1 + M_1\right)\left(E_2+M_2\right)}\Bigg( \bigg( 2 \sqrt{s} + M_1 + M_2 \bigg.\Bigg.
\\\nonumber
&-& \Bigg.\left. \left. (M_1-M_2)\frac{(m_2^2-m_1^2)}{m_{Vx}^2} \right) C_{t1, {\rm VB}}^{I} +\left(\frac{M_1+ M_2}{2 M}\left(2 \sqrt{s} + M_1 + M_2\right) + \frac{s-u}{2M}\right)C_{t2, {\rm VB}}^{I}\Bigg)\right]
\\\nonumber
&+&\left(C_{t1, {\rm VB}}^{I}+ \frac{M_1+ M_2}{2 M}C_{t2, {\rm VB}}^{I} \right)\left(-2 K_1\cdot \epsilon_2 \left[ \epsilon_1^0 - \frac{\vec \sigma\cdot\vec \epsilon_1 ~\vec \sigma\cdot\vec P_1}{E_1+M_1} - \frac{\vec \sigma\cdot\vec P_2 ~\vec \sigma\cdot\vec \epsilon_1}{E_2+M_2} + 
\frac{\epsilon_1^0 \vec \sigma\cdot\vec P_2 ~\vec \sigma\cdot\vec P_1}{\left(E_1 + M_1\right)\left(E_2+M_2\right)}\right]\right. \\\nonumber
&-&\left.2 K_2\cdot \epsilon_1 \left[ \epsilon_2^0 - \frac{\vec \sigma\cdot\vec \epsilon_2 ~\vec \sigma\cdot\vec P_1}{E_1+M_1} - \frac{\vec \sigma\cdot\vec P_2 ~\vec \sigma\cdot\vec \epsilon_2}{E_2+M_2} + 
\frac{\epsilon_2^0 \vec \sigma\cdot\vec P_2 ~\vec \sigma\cdot\vec P_1}{\left(E_1 + M_1\right)\left(E_2+M_2\right)}\right]\right) +\frac{2}{M}\left(K_1\cdot\epsilon_2 P_1\cdot\epsilon_1+K_2\cdot\epsilon_1 P_1\cdot\epsilon_2\right) \\\nonumber
&\times&\Biggl. \left(1  -  \frac{\vec \sigma\cdot\vec P_2 ~\vec \sigma\cdot\vec P_1}{\left(E_1 + M_1\right)\left(E_2+M_2\right)}  \right) C_{t2, {\rm VB}}^{I} \Biggr\} \sqrt{\frac{M_1 + E_1}{2 M_1}} \sqrt{\frac{M_2 + E_2}{2 M_2}},
\end{eqnarray}
\end{widetext}
where, $m_{Vx}$ represents the mass of the exchanged meson, $m_1(m_2)$, $K_1(K_2)$ represent the mass and four-momentum  of the meson in the initial (final) state, $M$ denotes the SU(3) average mass which is taken as the nucleon mass, $M_1 (M_2)$ and $E_1(E_2)$, $\vec P_1(\vec P_2)$ represent the mass, energy and three-momentum of the baryon in the initial (final) state. 
The values of $C^{I}_{t1, VB}$ and  $C^{I}_{t2, VB}$ are  given in Table~\ref{t1h}, \ref{t2h}, \ref{t13h}, \ref{t23h} in Appendix~\ref{coeff}. The s-wave projection of the  $t$-channel amplitude is also done numerically, as in the case of  the $u$-channel amplitude. We have followed the arguments of Ref.~\cite{raquelgeng} for the numerical integration of the $t$-channel amplitude.%However, it must be added here that for the sake of the numerical  integration, we write the denominator $1/(t-m_{Vx}^2) \rightarrow 1/(t-m_{Vx}^2 + i m_{Vx} \Gamma_{Vx})$, where $\Gamma_{Vx}$ is the width of the exchanged meson. This substitution can lead to an additional imaginary part in the poles eventually appearing in the complex plane which cannot be related to any physical process (as discussed in Ref.~\cite{raquelgeng}). Thus, to avoid the influence of this added imaginary part we take only the real part of the result obtained after the numerical integration, as also done in Ref.~\cite{raquelgeng}.} %The spin projected Eq.~(\ref{VtVB}) for total spin 1/2 and 3/2 gives
%\begin{eqnarray}\nonumber
%V^{I,1/2}_{t, {\rm VB}} &=&  \left(\frac{ |\vec{K}_1||\vec{K}_2|}{3m_1 m_2} - 1 \right)\frac{C_{t, {\rm VB}}^{I}}{4 f_V^2} \left(2 \sqrt{s} - M_1 - M_2 \right) \\
%&\times&\sqrt{\frac{M_1 + E_1}{2 M_1}} \sqrt{\frac{M_2 + E_2}{2 M_2}},
%-\frac{C^I_{t, VB}}{4 f_{V}^2}(\omega + \omega^\prime)  \vec{\epsilon}_1\cdot \vec{\epsilon}_2,
%\label{VtVBh}
%\end{eqnarray}
%and
%\begin{eqnarray}\nonumber
%V^{I,3/2}_{t, {\rm VB}} &=& \left(\frac{ 2|\vec{K}_1||\vec{K}_2|}{3m_1 m_2} - 1 \right) \frac{C_{t, {\rm VB}}^{I}}{4 f_V^2} \left(2 \sqrt{s} - M_1 - M_2 \right) \\
%&\times&\sqrt{\frac{M_1 + E_1}{2 M_1}} \sqrt{\frac{M_2 + E_2}{2 M_2}}.%-\frac{C^I_{t, VB}}{4 f_{V}^2}(\omega + \omega^\prime)  \vec{\epsilon}_1\cdot \vec{\epsilon}_2,
%\label{VtVB3h}
%\end{eqnarray}

\begin{figure}[h!]
\includegraphics[width=0.48\textwidth]{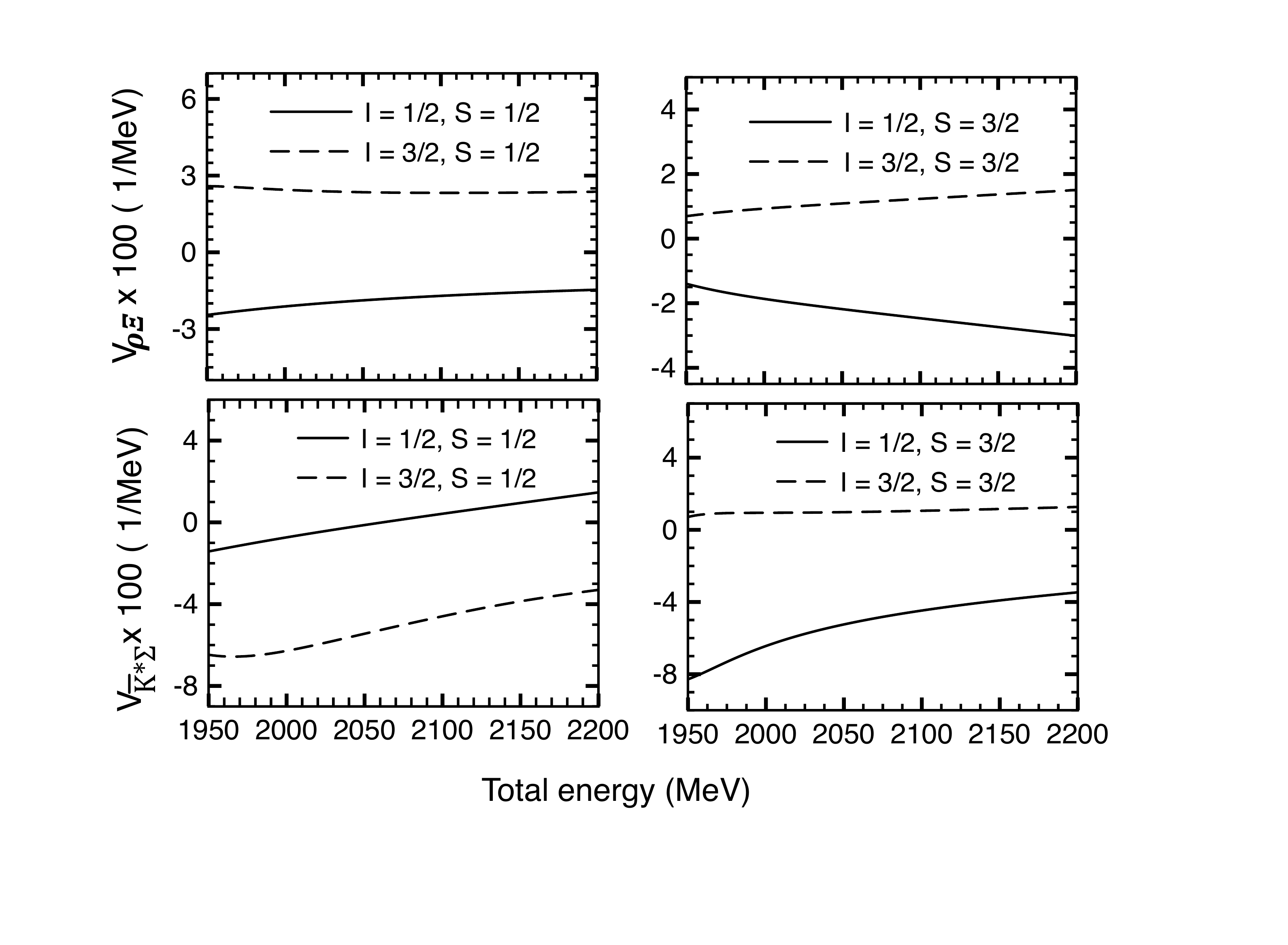}
\caption{A comparison of the lowest order $\rho \Xi$ and $\bar K^* \Sigma$ 
amplitudes  (Eq.~(\ref{VB})) for different spin-isospin configurations.  }\label{kernels}
\end{figure}

It can be seen that the amplitude of Eq.(\ref{Vtex}) is  spin degenerate at low energies (where only the $\vec \epsilon_1 \cdot \vec \epsilon_2$ spin structure contributes). This finding is in agreement with the results  found in Refs.~\cite{Kolomeitsev:2003kt,GarciaRecio:2003ks,ramosvb,Gamermann,Pavon}. However, such near degeneracy is removed by summing Eqs.~(\ref{contacts}), (\ref{u-channel}), (\ref{schannel3}) to the $t$-channel amplitude, giving rise to spin, isospin-dependent results. %Although, sometimes, it is possible that the interaction of a vector meson with a baryon may depend weakly on spin configurations, we find that this is not always the case.  For example, the contact-term (Eq.~(\ref{contacts})) for the $\rho \Xi$ diagonal amplitude, or for the transition of $\bar K^*\Lambda$ to $\rho \Xi$, $\omega \Xi$, $\phi \Xi$, is small. But the  same contact term is comparable to, or even larger than, the $t$-channel amplitude for $\bar K^* \Sigma \to \bar K^* \Sigma, \bar K^*\Lambda, \rho \Xi, \omega \Xi, \phi \Xi$. 
For a numerical comparison, we show the lowest order amplitudes obtained in the present work for the $\rho \Xi$ and $\bar K^* \Sigma$ channels, as examples, in Fig.~\ref{kernels}.%and $\bar K^* \Sigma$ channels, as examples, in Fig.~\ref{kernels}. 
In fact, the general spin-dependence of the vector meson-baryon kernels may be expected since both the interacting hadrons have nonzero spin and similar masses.  An alternative mechanism of breaking the spin-degeneracy has been suggested in Ref.~\cite{javi}. Further, we go a step forward by including pseudoscalar-baryon as coupled channels in our formalism. The consideration of various contributions to the vector-baryon amplitudes, together with the treatment of the vector- and pseudoscalar-mesons at par in the coupled channel dynamics, is the distinct feature of our formalism.

A formalism to obtain the transition amplitudes  between the pseudoscalar-baryon and the vector-baryon channels was developed in our previous work \cite{pbvb}, where the Kroll-Ruderman term for the photoproduction of a pion was modified, in consistency with
 the vector meson dominance phenomenon, by replacing the photon by a vector meson. The deduction was extended to the  SU(3) case in Ref.~\cite{pbvb} to obtain a general Lagrangian for the transitions among pseudoscalar-baryon (PB) and  vector-baryon (VB) channels 
 \begin{eqnarray} \label{pbvbeq}
\mathcal{L}_{\rm PBVB} &=& \frac{-i g_{KR}}{2 f_\pi} \left( F^\prime \langle \bar{B} \gamma_\mu \gamma_5 \left[ \left[ P, V^\mu \right], B \right] \rangle \right.\\\nonumber &+& 
\left. D^\prime \langle \bar{B} \gamma_\mu \gamma_5 \left\{ \left[ P, V^\mu \right], B \right\}  \rangle \right),
\end{eqnarray}
where  $F^\prime = 0.46$, $D^\prime = 0.8$ reproduce the axial coupling constant of the nucleon:  $F^\prime + D^\prime \simeq  g_A = 1.26$ \cite{pbvb}.

In the present work we need the transition amplitudes between pseudoscalar-baryon and vector-baryon channels with strangeness $-2$. We obtain them using Eq.~(\ref{pbvbeq})
%\begin{equation}
%V^I_{\textrm PBVB}=- i \,C^I_{\rm PB VB} \frac{g_{\textrm{KR}}}{2\sqrt{f_P f_V}}  \vec\sigma\cdot\vec\epsilon,\label{KR}
%\end{equation}
{\small\begin{eqnarray}\nonumber
V^{I}_{\textrm PBVB} &=& - i C_{\rm PBVB}^{I} \frac{g_{\textrm{KR}}}{2 \sqrt{f_P f_V}} \left\{ \epsilon^0 \left( \frac{\vec{\sigma}\cdot\vec{P}_1}{E_1 + M_1} + \frac{\vec{\sigma}\cdot\vec{P}_2}{E_2 + M_2}\right) \right.\\\nonumber
&-& \left.\vec{\sigma}\cdot\vec{\epsilon} -  \frac{\vec{\sigma}\cdot\vec{P}_2~\vec{\sigma}\cdot\vec{\epsilon}~\vec{\sigma}\cdot\vec{P}_1}{\left(E_1 + M_1\right) \left( E_2 + M_2 \right)}\right\}\sqrt{\frac{M_1 + E_1}{2 M_1}} \\
&\times&\sqrt{\frac{M_2 + E_2}{2 M_2}},
\end{eqnarray}}
which, on s-wave projection, gives
{\small \begin{eqnarray}\nonumber
V^{I,S}_{\textrm PBVB} &=& - i C_{\rm PBVB}^{I} \frac{g_{\textrm{KR}}}{2 \sqrt{f_P f_V}} \Biggl\{\!\! - \vec{\sigma}\cdot\vec{\epsilon}+ \epsilon^0\! \left(\!\! \frac{\vec{\sigma}\cdot\vec{P}_1}{E_1 + M_1} \!\!\Biggr)\!\! \right\}\!\!\\\label{KR}
&\times&\sqrt{\frac{M_1 + E_1}{2 M_1}} \sqrt{\frac{M_2 + E_2}{2 M_2}}
\end{eqnarray}}
\noindent
In Eqs.~(\ref{pbvbeq})-(\ref{KR}) $g_{\textrm{KR}}$ is the Kroll-Ruderman coupling~\cite{pbvb,hyperons,more}
\begin{equation}
g_{\textrm{KR}}=\frac{m_V}{ \sqrt{2 f_P f_{V}}}, \label{gkr}
\end{equation}
with $m_V$ denoting the mass of the vector meson and $f_P$ and $f_V$ being the decay constants of the pseudoscalar and vector meson. The values of the isospin coefficients $C^I_{\rm PB VB}$ for isospin 1/2 and 3/2 are listed in Table~\ref{kr_iso1/2}~and~\ref{kr_iso3/2} in the Appendix~\ref{coeff}, respectively. The  PB $\leftrightarrow$ VB transition amplitudes  for spin 1/2 and 3/2 are obtained as
{\small
\begin{eqnarray}\label{KR1}
&V^{I,1/2}_{\textrm PBVB}&  =\!  iC_{\rm PBVB}^{I}\! \frac{g_{\rm KR}}{2 \sqrt{f_P f_V}}\!\! \left\{\!\!\sqrt{3}\! +\! \sqrt{\frac{1}{3}}\!\frac{\mid\!\vec{K}_v\!\mid}{m_v} \frac{\mid \vec{P}_1\mid}{\left( E_1 + M_1 \right)}\!\!\right\}\!,\!\\
&V^{I,3/2}_{\textrm PBVB}&= - i C_{\rm PBVB}^{I} \frac{g_{\rm KR}}{2 \sqrt{f_P f_V}} \sqrt{\frac{2}{3}} \frac{\mid \vec{K}_v \mid}{m_v}   \frac{\mid \vec{P}_1 \mid} {\left( E_1 + M_1 \right)}.\label{KR2}
\end{eqnarray}} %for both isospin configurations are  spin $1/2$ amplitudes
We find that the PB $\leftrightarrow$ VB amplitudes give negligible contribution in the spin 3/2 case, as can be seen from the kinematic dependence in Eq~(\ref{KR2}). 
%The  PB $\leftrightarrow$ VB transitions  for total spin 3/2 is zero (as in Refs.~\cite{pbvb,hyperons,more}). 
This is consistent with the results obtained within a different formalism~\cite{javi}, where the VB amplitudes in spin 3/2 have been found to change weakly when coupled  to pseudoscalar baryon  systems.

Finally, the pseudoscalar meson-baryon amplitudes are obtained from the lowest order chiral Lagrangian
%\begin{equation}
%\mathcal{L_{PB}} = \langle \bar B i \gamma^\mu \bigtriangledown_\mu B \rangle - M_B \langle \bar B B \rangle + \frac{1}{2} D \langle \bar B i \gamma^\mu \bigtriangledown_\mu B \rangle,
%\end{equation}
\cite{ecker,pich}
\begin{eqnarray}
\mathcal{L}_{PB} &=& \langle \bar B i \gamma^\mu \partial_\mu B  + \bar B i \gamma^\mu[ \Gamma_\mu, B] \rangle - M_{B} \langle \bar B B \rangle  \\\nonumber
&+&  \frac{1}{2} D^\prime \langle \bar B \gamma^\mu \gamma_5 \{ u_\mu, B \} \rangle + \frac{1}{2} F^\prime \langle \bar B \gamma^\mu \gamma_5 [ u_\mu, B ] \rangle,\label{LPB}
\end{eqnarray}
where  
\begin{eqnarray}\nonumber
\Gamma_\mu &=& \frac{1}{2} \left( u^\dagger \partial_\mu u + u \partial_\mu u^\dagger  \right), \,u_\mu = i u^\dagger \partial_\mu U u^\dagger ,\,\\
U&=&u^2 = {\textrm exp} \left(i \frac{P}{f_P}\right).\label{gammau}
\end{eqnarray}
As defined earlier, $f_P$ in Eq.~(\ref{gammau}) is the pseudoscalar decay constant, and  $P$ is
\begin{eqnarray} \nonumber
P &=&
\left( \begin{array}{ccc}
\pi^0 + \frac{1}{\sqrt{3}}\eta & \sqrt{2}\pi^+ & \sqrt{2}K^{+}\\
&& \\
\sqrt{2}\pi^-& -\pi^0 + \frac{1}{\sqrt{3}}\eta & \sqrt{2}K^{0}\\
&&\\
\sqrt{2}K^{-} &\sqrt{2}\bar{K}^{0} & \frac{-2 }{\sqrt{3}} \eta
\end{array}\right)
\end{eqnarray}
The resulting s-wave amplitudes are consistent with those obtained earlier in Ref.~\cite{ramosXi},
{\small \begin{equation}
V^{I,1/2}_{PB} = - \frac{C^I_{PB}}{4 f_1 f_2} \left(2 \sqrt{s} - M_1 - M_2 \right) \sqrt{\frac{\left(M_1 + E_1 \right)\left(M_2 + E_2 \right)}{4 M_1 M_2}},
\end{equation}}
where  $M_1$ ($M_2$),  $E_1$ ($E_2$) represent the mass and energy of the baryon in the initial (final) state, respectively, and $f_1$ ($f_2$) is the decay constant of the meson in the initial (final) state. We do not give the values of 
$C^I_{PB}$ here since they are same as those given in Ref.~\cite{ramosXi}.
 
With the kernels prepared we solve the Bethe-Salpeter equation
 in its on-shell factorization form ~\cite{osetramos,oller}
\begin{align}
T=(1-VG)^{-1} V,\label{BS}
\end{align}
where $G$ is a loop of two hadrons which is divergent in nature. It is possible to evaluate the loop function by using the dimensional regularization or with a three-momentum cutoff. In the former case, the subtraction constants can be fixed to have natural values following Refs.~\cite{Gamermann,hyodo,Oller:2000fj}, to ensure the interpretation of the resonances found in the calculations as those arising from the dynamics of the system. In the latter case, a cutoff parameter $\Lambda$ is used to regularize the loop function using a Gaussian form factor as 
 \begin{align}
 G_i(\sqrt{s})\!\! = \!\! \int\limits_0^\infty \frac{d^3q}{\left(2 \pi \right)^3} \frac{1}{2 E_{1i} \left(\vec{q}\right)} \frac{2 M_i}{2 E_{2i} \left(\vec{q}\right)} \dfrac{e^{\left[{-\left(q^2 - q_{i,on}^2 \right)/\Lambda^2}\right]}}{\sqrt{s} - E_{1i} \left(\vec{q} \right) - E_{2i} \left(\vec{q} \right)},\label{loopcutoff}
 \end{align} 
where the subscript $i$ is the label for the $i$th-channel,  $E_{1i}$ and $E_{2i}$ represent the energies of the two particles in the channel, $M_i$ is the mass of the propagating baryon and, $q_{i,on}$  is the three momentum of the relative motion of the particles in the intermediate state when they are on mass shell. We have calculated the amplitudes using both methods and find that the results obtained and conclusions are very similar. In the following we show the results obtained by calculating the loop function using Eq.~(\ref{loopcutoff}) with $\Lambda = 800$~MeV, although, later on, we will present the values of the poles found within the dimensional regularization scheme  too. 

It is important to add here that we take into account the fact that the width of some vector mesons, like $K^*$ and $\rho$, is considerably large. To do this we convolute  the loops over the varied mass of these mesons following  Ref.~\cite{ramosvb}, 
{\small \begin{eqnarray}\nonumber
\tilde G_i (\sqrt{s})\!\!&=&\!\dfrac{1}{N}\! \int\limits_{(m-2\Gamma)^2}^{(m+2\Gamma)^2}\!\!\!\!\!\!\!\! d\tilde{m}^2\!\! \left(\!\!\dfrac{-1 }{\pi}\!\!\right) G_i(\sqrt{s})\!\! \\
&\times&{\textrm Im} \left\{\!\!\dfrac{1}{\tilde{m}^2 - m^2 +  i m\Gamma(\tilde{m})}\!\!\right\}\!\!,\label{Gconv}
\end{eqnarray}}
where $G_i(\sqrt{s})$ is calculated using Eq.~(\ref{loopcutoff}) and where
\begin{eqnarray}\nonumber
N = \int\limits_{(m-2\Gamma)^2}^{(m+2\Gamma)^2}  d\tilde{m}^2 \left( - \dfrac{1}{\pi}\right) {\rm Im} \left\{\dfrac{1}{\tilde{m}^2 - m^2 + i m \Gamma(\tilde{m})}\right\},
\end{eqnarray}
with
{\small
\begin{equation}\nonumber
\Gamma(\tilde{m})\!\!=\!\!\Gamma_{\rm meson}\!\! \left(\!\dfrac{m^2}{\tilde{m}^2}\!\right)\!\!\!\!  \left(\dfrac{\lambda^{1/2}(\tilde{m}^2, m_d^2, m_d^{\prime\, 2})/2\tilde{m}} {\lambda^{1/2}(m^2, m_d^2, m_d^{\prime\, 2})/2m}\right)^3\!\!\!\!  \theta\left( \tilde{m} - m_d - m_d^\prime \right)\!.
\end{equation}}
\noindent
In the above equation, $m_d, m_d^\prime$ denote the masses of the decay products of the vector mesons,  i.e., pion masses in case of $\rho$, and
kaon and pion mass in case of $K^*$.  We use $\Gamma_\rho$ and $\Gamma_{K^*} $ as 149.4~MeV and 50.5~MeV, respectively.   Thus, for the channels involving the $\rho$ and  the $K^*$ mesons, we use Eq.~(\ref{Gconv}) to solve  the Bethe-Salpeter equation (Eq.~(\ref{BS})).

To search for resonances/bound states we need to obtain the $T$-matrix of  Eq.~(\ref{BS}) in the complex energy plane, for which we calculate the loop function in the first (I) and second (II) Riemann sheet as \cite{oller,inoue}: 
\begin{eqnarray}\nonumber
G_i(\sqrt{s}) = \left\{\begin{array}{cc}
G_i^{(I)}(\sqrt{s}), & {\rm for }~ {\rm Re}\{\sqrt{s}\}\!<\!(m_i\!+\!M_i)\\
&\\
G_i^{(II)}(\sqrt{s}), & {\rm for }~ {\rm Re}\{\sqrt{s}\} \ge \!(m_i\!+\!M_i)
\end{array},\right.
\end{eqnarray}
where
\begin{eqnarray}\label{ana_contG1}
G_i^{(I)}(\sqrt{s})&=& G_i(\sqrt{s}) \\\nonumber
G_i^{(II)}(\sqrt{s})&=& G_i^{(I)}(\sqrt{s}) -2\,i\,{\rm Im} \{G_i^{I}\} \\
&=& G_i^{(I)}(\sqrt{s}) + i \frac{M_i q}{2\pi \sqrt{s}}, \label{ana_contG2}
\end{eqnarray}
with $(m_i+M_i)$ being the threshold of the $i$th-channel.  If a pole appears in the complex plane, it can be seen in the complex amplitude for all the channels. Depending on the threshold of a given channel, the pole can appear below or above the threshold (i.e, on the corresponding first or second Riemann sheet of that channel).

\section{Results and discussions}
\subsection{Isospin = 1/2, Spin =1/2 and Parity = $-1$}
\begin{figure}[h!]
\includegraphics[width=0.51\textwidth]{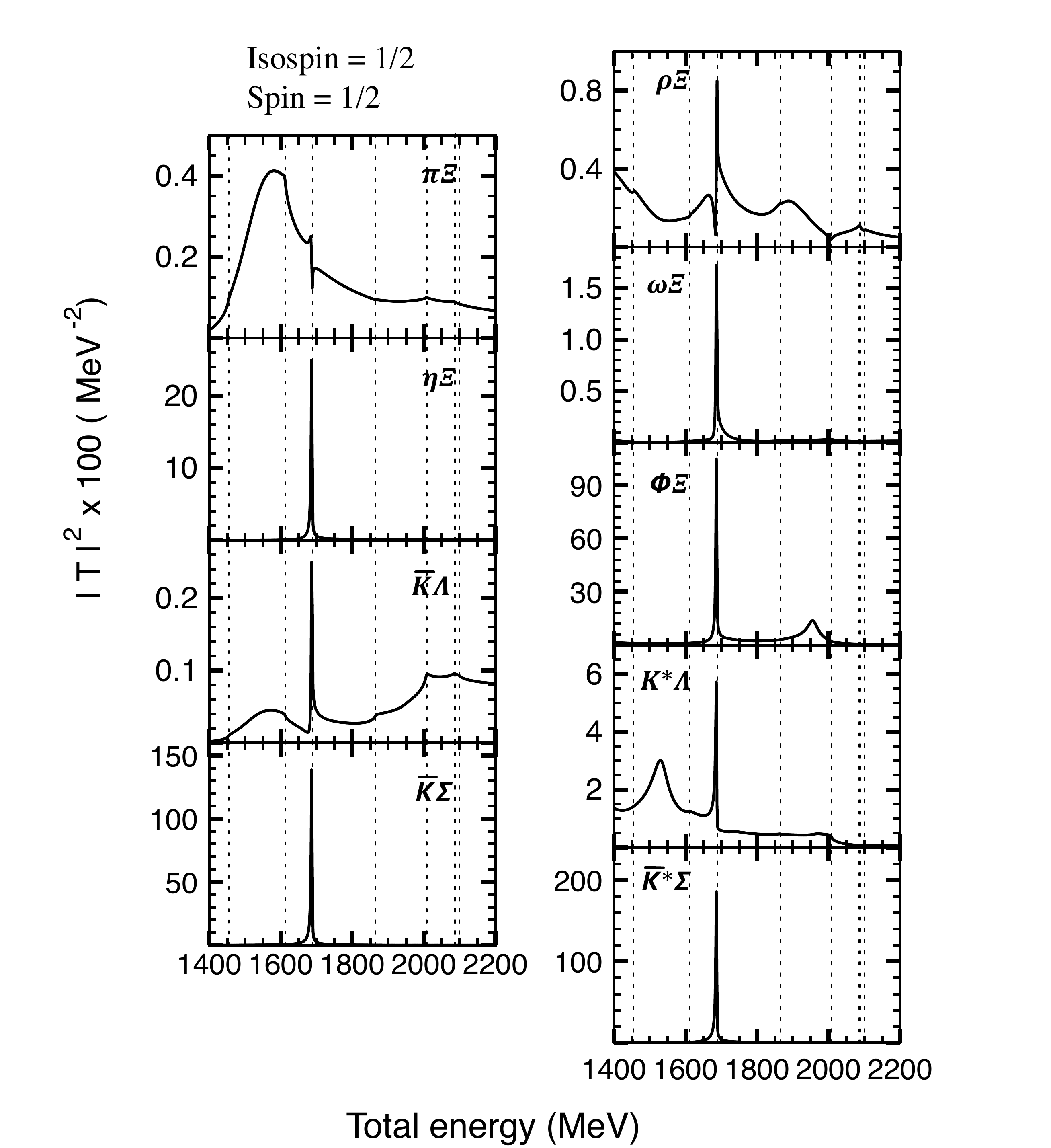}
\caption{Squared amplitudes for the total isospin and spin 1/2. The vertical dashed lines indicate the thresholds as summarized in Table~\ref{masses}.  }\label{ampl_Sh}
\end{figure}

\begin{figure}[h!]
\includegraphics[width=0.41\textwidth]{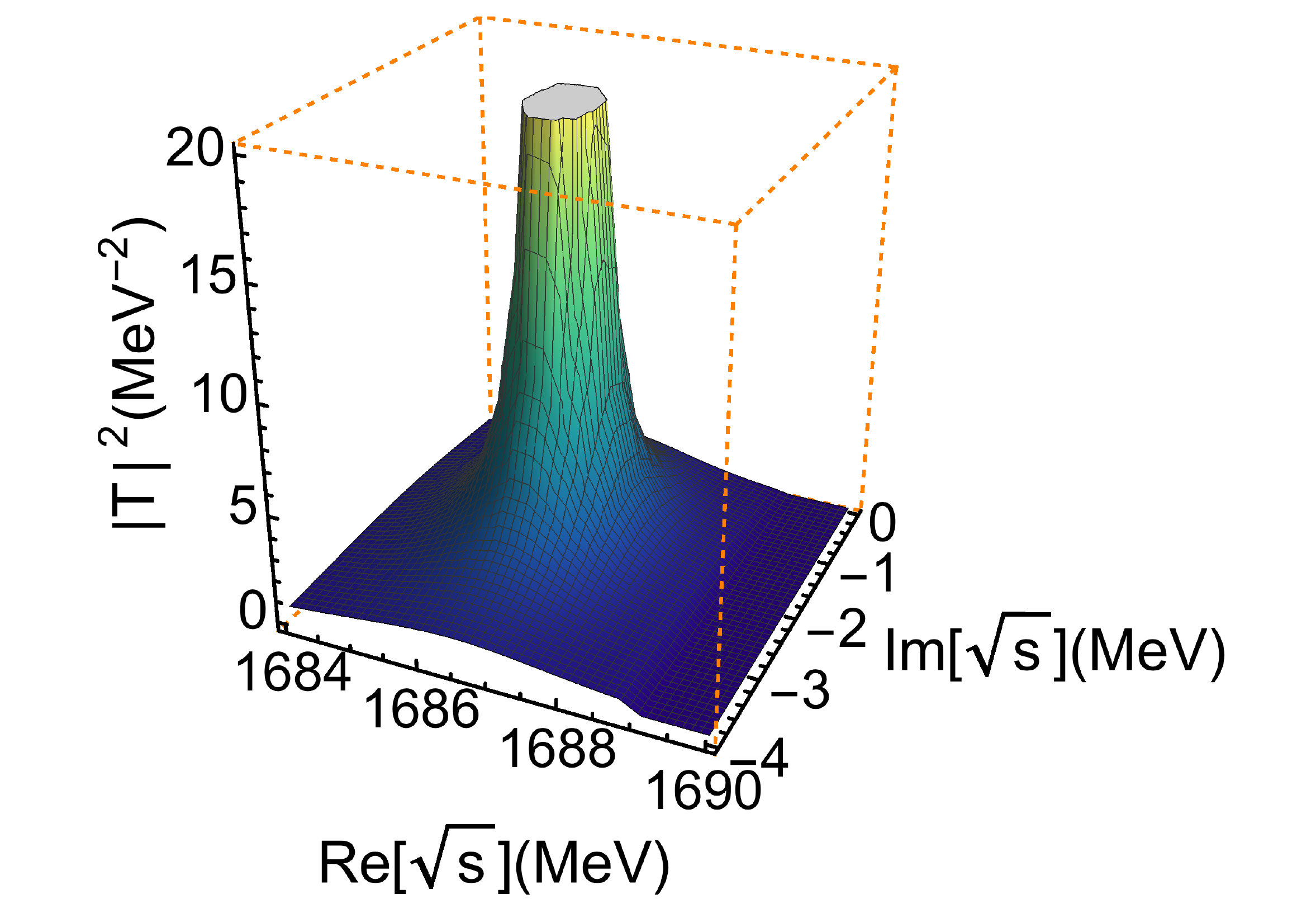}
\caption{Squared amplitudes for the $\bar K \Sigma$ channel, in the complex plane, with total isospin and spin 1/2.  }\label{fig:pole1690}
\end{figure}

We begin the discussion of the results by showing the squared amplitudes for different coupled channels in the isospin 1/2, spin 1/2 configuration in Fig.~\ref{ampl_Sh}.
It can be seen from Fig.~\ref{ampl_Sh} that a clear narrow peak is present around 1690 MeV in all the channels except in $\pi \Xi$. This peak corresponds to a pole in the complex plane (shown in Fig.~\ref{fig:pole1690}) at 
\begin{align}\label{M1690}
M - i \Gamma/2 = 1687 - i 2 ~{\rm MeV}.
\end{align} 
This pole position is in good agreement with the recently determined mass and width of $\Xi(1690)$ by the {\it BABAR} Collaboration: $M~=~\left(1684.7 \pm 1.3^{+2.2}_{-1.6} \right)$~MeV, $\Gamma = \left(8.1^{+3.9+1.0}_{-3.5-0.9}\right)$ MeV \cite{babar} and those determined by the BELLE Collaboration:
  $M  = \left(1688 \pm 2\right)$ MeV, $\Gamma = \left(11 \pm 4\right)$ MeV \cite{belle}. The spin-parity $1/2^-$ also coincides with the experimental data analysis of Ref.~\cite{babar}. The full width of 4 MeV in Eq.~(\ref{M1690}) is a bit larger than the value $\sim 1$ MeV  obtained in Ref.~\cite{Sekihara}. This is because the coupling of the vector-baryon channels pushes the pole to slightly lower energies, hence, bringing it slightly closer to the $\bar K \Lambda$ channel. This latter fact allows for a bigger decay width since the coupling of the pole to the $\bar K \Lambda$ channel increases. Another finding is that in our work, where we insist on using the cutoff variable related to the natural subtraction constants such that the resulting poles can be related to dynamically generated states \cite{Gamermann,hyodo,Oller:2000fj}, we do not find the pole for $\Xi(1690)$ if we switch off the coupling between PB and VB channels.

We have also calculated the branching ratios (in percentage) of our state to  the different open channels for decay 
\begin{equation}\nonumber
\frac{\Gamma_{R \to i}}{\sum\limits_i \Gamma_{R \to i}},
\end{equation}
where $\Gamma_{R \to i}$ represents the partial decay width of the resonance, $R$, to the $i$th  channel, and the sum in the denominator, in case of $\Xi(1690)$ is
\begin{equation}\nonumber
\sum\limits_i \Gamma_{R \to i} = \Gamma_{\Xi(1690) \to \bar K \Lambda} +\Gamma_{\Xi(1690) \to \bar K \Sigma} +\Gamma_{\Xi(1690) \to \pi \Xi} 
\end{equation}
The partial decay width is obtained as \cite{oller}, 
\begin{eqnarray}\label{dec_rate}
\Gamma_{R\to 1+2}&=&  \int\limits_{W_{\rm min}}^{W_{\rm max}} \frac{dW}{W^2} \frac{ q} {4 \pi^2} M_R M_2 \mid g_{R12} \mid^2 \\\nonumber
&\times&\left[\frac{\Gamma_R}{(M_R - W)^2 + (\Gamma_R/2)^2}\right],
\end{eqnarray}
 where $M_R$, $\Gamma_R$, $M_2$ denote the mass and width of the decaying resonance and the mass of the baryon in the final state,  $q$ represents the magnitude of the relative momentum for the final meson-baryon in the center of mass frame, and $g_{R12}$ represents the coupling of the resonance to the decay channel.   The coupling of the different channels to the state we associate with $\Xi(1690)$ are given in Table.~\ref{coupXi1690}. The variable mass of the resonance, $W$, in Eq.~(\ref{dec_rate}) runs from the minimum value, $W_{\rm min}$, which can be as low as the threshold of the decay channel, to a maximum value, $W_{\rm max}$, above the resonance mass, which can be as large as $\infty$. However, the precise value of these limits of integration are taken such that  
 \begin{equation}
 \Gamma_{R, {\rm tot}} = \sum_i \Gamma_{R \to i},\label{Gamma_sum}
 \end{equation}
where $\Gamma_{R, {\rm tot}}$ is   the total width of the resonance obtained from the pole position. %and $\Gamma_{R, i}$ is the partial width to the $i$th  channel open for decay. 
In the present case of $\Xi(1690)$, it turns out that using $W_{\rm min} = M_R - 1.6 \Gamma_{R, {\rm tot}}$ (if $W_{\rm min} >$ threshold of the decay channel) and  $W_{\rm max} = M_R + 1.6 \Gamma_{R, {\rm tot}}$ in Eq.~(\ref{dec_rate}) satisfies the condition in Eq.~(\ref{Gamma_sum}). Note that  these values of $W_{\rm min}$ and $W_{\rm max}$,   coincidentally,  happen to be same as those used in Ref.~\cite{oller} for $f_0(980) \to \pi \pi,~ K \bar K$.
 
The branching ratios of our state to $\pi \Xi$, $\bar K \Lambda$ and $\bar K \Sigma$ are obtained as 17$\%$, $28.5\%$ and $54.5\%$, respectively. Using these values, we can see that the branching fraction defined by Eq.~(\ref{ratio}) is 0.52, which is remarkably similar to the experimental value \cite{belle}.

The coupling of  a resonance to the different channels can be calculated by recalling that the scattering matrix for the transition $i \to j$, $T_{ij}$, in the vicinity of a pole can be written as
\begin{equation}
T_{ij} = \frac{g_i g_j}{z - z_R},
\end{equation}
where $z_R$ corresponds to the pole position associated to the resonance in the complex plane. Then, the product of the couplings $g_i g_j$ can be identified with the residue, $R_{ij}$, of $T_{ij}$, which can be calculated via an integration of $T_{ij}$ along a closed contour around pole. In this way, $g_i^2 = R_{ii}$, and  we can write
\begin{equation}
g_j = \frac{R_{ij}}{\sqrt{R_{ii}}}.
\end{equation}
\begin{table}[h!]
\caption[]{ Couplings of different channels to $\Xi(1690)$.  
} \label{coupXi1690}
\begin{ruledtabular}
\begin{tabular}{cc}
Channels &Couplings   \\
\hline
$\pi \Xi$              &$-0.1 - i 0.1$ \\
$\eta \Xi$             &$~0.9 + i 0.2$ \\
$\bar K \Sigma$       & $~1.5 + i 0.2$\\
$\bar K \Lambda$          &$-0.3 + i 0.1$  \\
$\rho \Xi$           &$~0.2 - i 0.3$ \\
$\omega \Xi$&$-0.4 + i 0.2$ \\
$\phi \Xi$          &$~1.2 + i 0.5$ \\
$\bar K^* \Sigma$       &$-1.4 - i 0.4$ \\
$\bar K^*\Lambda$        &$~0.6 + i 0.0$ \\
\end{tabular}
\end{ruledtabular}
\end{table}
It can be seen from the values of these couplings that the pole given by Eq.~(\ref{M1690}) couples very weakly to $\pi \Xi$, thus, explaining the difficulty of identifying the state in the data on $\pi  \Xi$ mass spectrum~\cite{babar}. The small coupling of the state to  $\bar K \Lambda$ and the stronger one to $\bar K \Sigma$, in spite of  the presence of a larger phase space for decay to the former 
(the masses and the thresholds of different channels are listed in Table.~\ref{masses})
\begin{table}[h!]
\caption[]{The isospin averaged masses of different particles and the thresholds of different channels. These thresholds are indicated in Figs.~\ref{ampl_Sh}~and~\ref{ampl_S3h} as vertical dashed lines.
} \label{masses}
\begin{ruledtabular}
\begin{tabular}{cc}
Channels  & Thresholds (MeV)  \\
(Masses are given in brackets, in MeV)& \\
\hline
$\pi (137) \Xi (1318)$                   &1455 \\
$\eta(547) \Xi(1318)$                  &1865 \\
$\bar K(496) \Sigma(1193) $       & 1689\\
$\bar K (496) \Lambda(1116)$          &1612  \\
$\rho (770) \Xi(1318)$                   & 2088\\
$\omega (782) \Xi(1318)$          &2100 \\
$\phi(1020) \Xi(1318)$          &2338 \\
$\bar K^*(892) \Sigma(1193) $       & 2085\\
$\bar K^*(892) \Lambda(1116)$        & 2008\\
\end{tabular}
\end{ruledtabular}
\end{table}
explains why the branching ratio given in Eq.~(\ref{ratio}) is larger than expected.  Thus, all the findings related to $\Xi(1690)$ can be well explained within the coupled channel dynamics of meson-baryon systems which indicate that this state should be interpreted as a dynamically generated state. It should be mentioned that suggestions on the hadronic dynamical origin of $\Xi(1690)$ have also been made in  Refs.~\cite{Sekihara, Kolomeitsev:2003kt,Gamermann,Oh}.

\begin{figure}[h!] 
\includegraphics[width=0.4\textwidth]{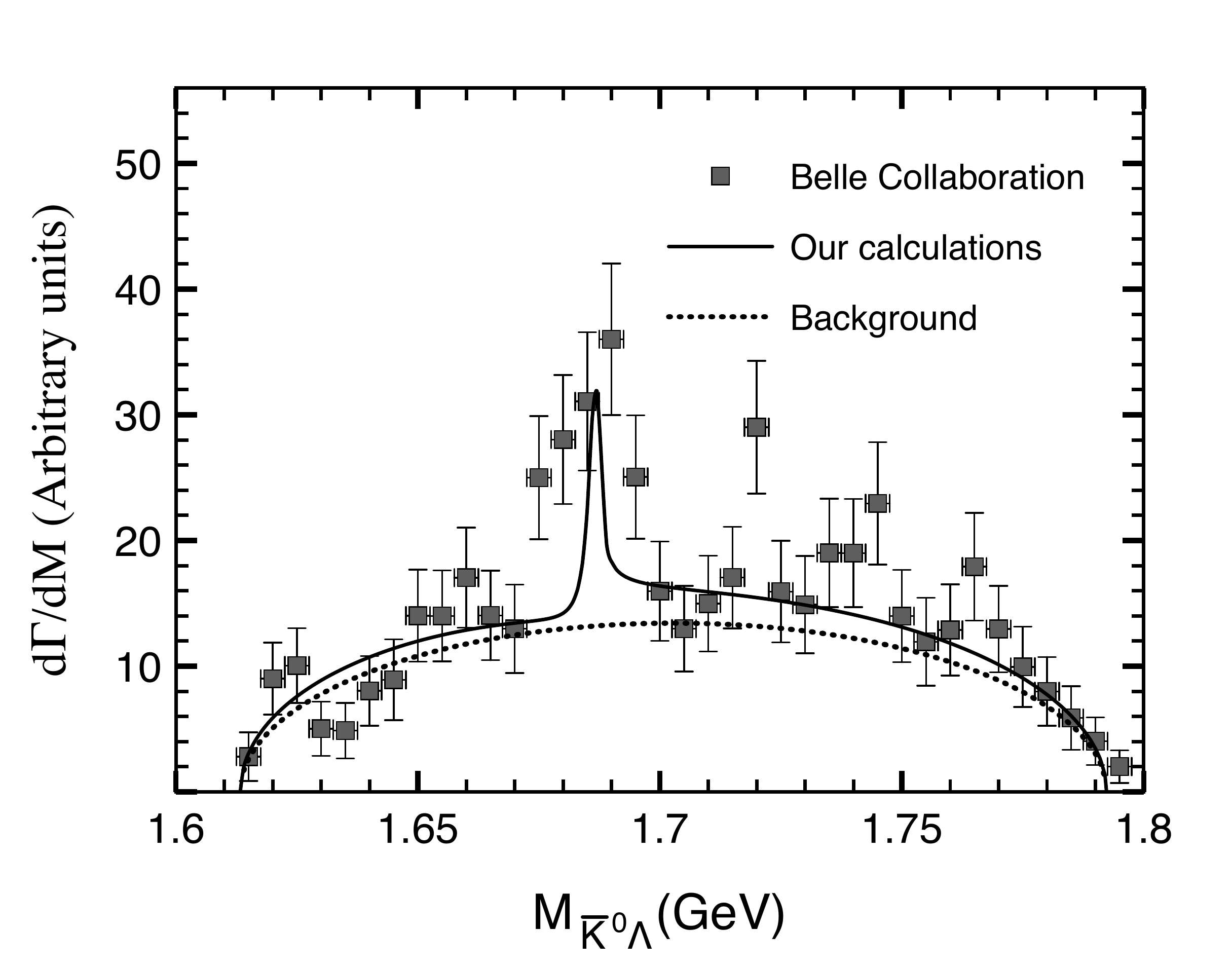}
\caption{$\bar K^0 \Lambda$ invariant mass spectra for the process $\Lambda_c \to K^+ \bar K^0 \Lambda^0$. The data is taken from Ref.~\cite{belle}.}\label{mass_spec1}
\end{figure}

\begin{figure}[h!]
\includegraphics[width=0.4\textwidth]{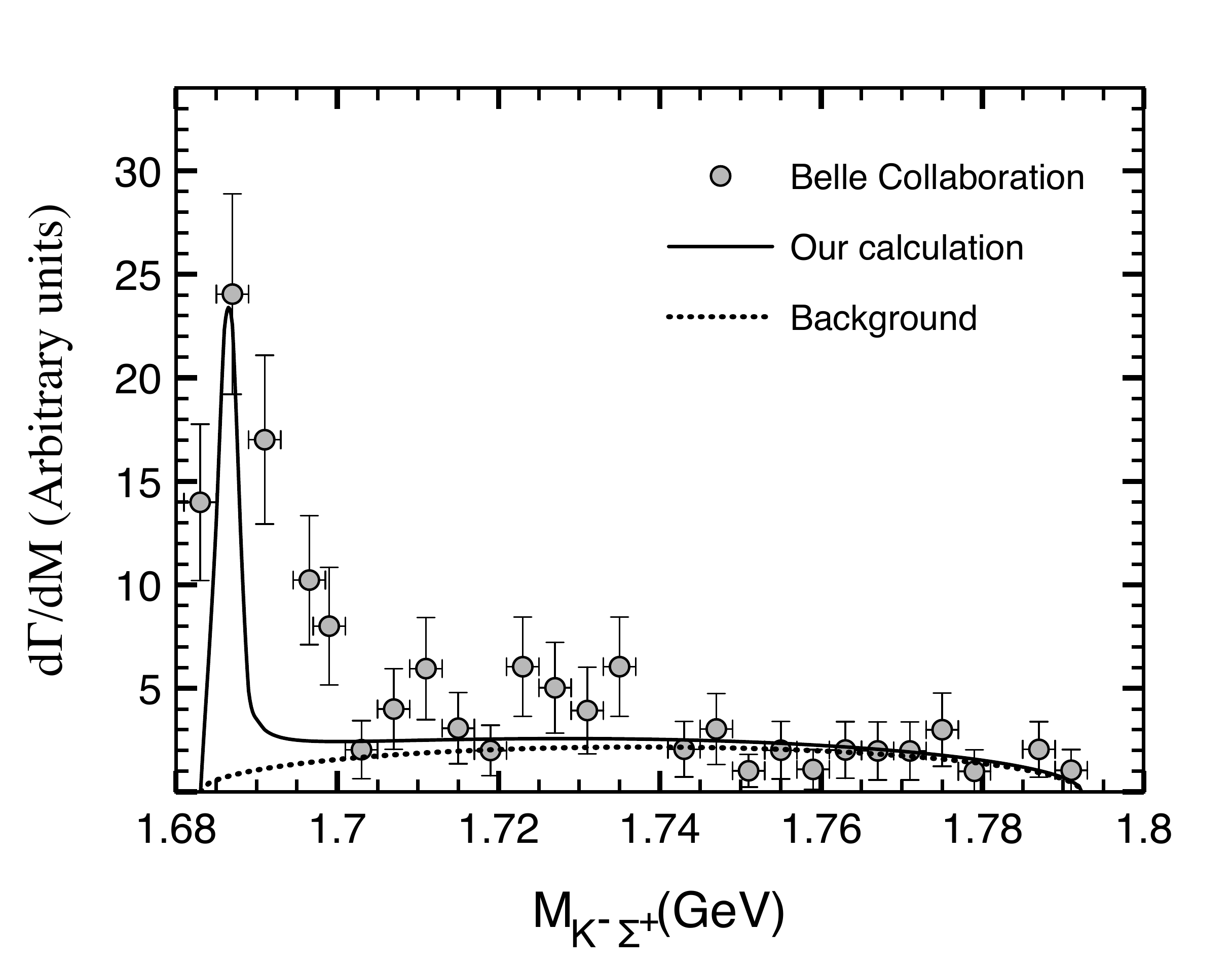}
\caption{$K^- \Sigma^+$ invariant mass spectra for the process $\Lambda_c \to K^+ K^- \Sigma^+$. The data is taken from Ref.~\cite{belle}.}\label{mass_spec2}
\end{figure}

Further, we find it useful to obtain the $\bar K \Lambda$ and $\bar K \Sigma$ invariant mass spectra and compare with the corresponding data available from the processes $\Lambda_c \to K^+ \bar K^0 \Lambda^0$ and $\Lambda_c \to K^+ K^- \Sigma^+$  in Ref.~\cite{belle}. The mass spectra are calculated using  
\begin{equation}
\frac{d \Gamma}{d M_{\bar K B}} \propto  P^*_{K^+} P^*_{\bar K B} |T|^2,\label{eq:mspec}
\end{equation}
%\frac{1}{\left(2 \pi \right)^3} \frac{M_B}{2 M_{\Lambda_c}}
where the $\bar K B$ subscript represents the $\bar K^0 \Lambda^0$ or $K^- \Sigma^+$ system, $P^*_{K^+}$ is the  momentum of $K^+$ in the $\Lambda_c$ rest frame and $P^*_{\bar K B}$ is the center of mass momentum in the $\bar K B$ rest frame. Further, to compare with the data outside the peak region, we add  to the $\bar K B$ $T$-matrix  a small background proportional to the phase space. To be explicit, $T$ in Eq.~(\ref{eq:mspec}) is 
\begin{equation}
T = T_{\bar K B} + \tilde B,
\end{equation}
where $\tilde B = 2 \times 10^{-3}$ MeV$^{-1}$ for  the $\bar K^0 \Lambda$ mass spectrum  and $\tilde B = 5 \times 10^{-2}$ MeV$^{-1}$ for the $\bar K^+ \Sigma^-$ case. A comparison of our results (which is multiplied by an arbitrary constant factor), as well as the background contribution, with the data are shown in Figs~\ref{mass_spec1} and \ref{mass_spec2}. 

 A technical remark is here in order.  To compare our results with the data, we calculate the momenta in Eq.~(\ref{eq:mspec}) using the physical masses of the mesons and baryons involved in the process, although the  $|T|^2$  has been calculated using average masses (as mentioned earlier). We would like to add here that if we would have used isospin average masses in the phase space to obtain the invariant mass distribution, a peak structure, though closer to the average $\bar K \Sigma$ threshold, would appear in Fig.~\ref{mass_spec2}  in spite of having the pole (given by Eq.~(\ref{M1690})) below the average $\bar K \Sigma$ threshold.  The latter would be a consequence of  the finite width of the pole (= 4 MeV). A manifestation of the resonance with the mass below the threshold of a particular channel but with a finite width, allowing the tail of the resonance fall in the physical energy region,  has been discussed in Ref.~\cite{MartinezTorres:2012du}. 

We have also solved the Bethe-Salpeter equation by calculating the loop function within the dimensional regularization scheme by using natural values of the subtraction constants:
\begin{equation}
\begin{array}{ccc}
a_{\pi \Xi} = -1.05 & a_{\eta \Xi} = -2.30 & a_{\bar K \Sigma} = -1.90 \\
a_{\bar K \Lambda} = -1.65 & a_{\rho \Xi} = -2.63 & a_{\omega \Xi} = -2.64 \\
a_{\phi \Xi} = -2.93 & a_{\bar K^* \Sigma} = -2.60&a_{\bar K^* \Lambda} = -2.50.\\
\end{array}
\end{equation}
These constants have been obtained by using the condition 
\begin{equation}
G [\mu =630~{\rm MeV}, \; \sqrt{s} = \sqrt{s}_{thr, min} ] = 0,
\end{equation}
for all the channels, where $\sqrt{s}_{thr, min}$ is the lowest threshold. 
Such a calculation leads to a pole at $1688 - i 1$ MeV, which is similar to Eq.~(\ref{M1690}). All other results obtained in our work are comparable in a similar way when obtained by using a cutoff  or dimensional regularization to calculate the loop integrals.  We, thus, would show only the results obtained with a cutoff in the following discussions.

\begin{figure}[h!]
\includegraphics[width=0.4\textwidth]{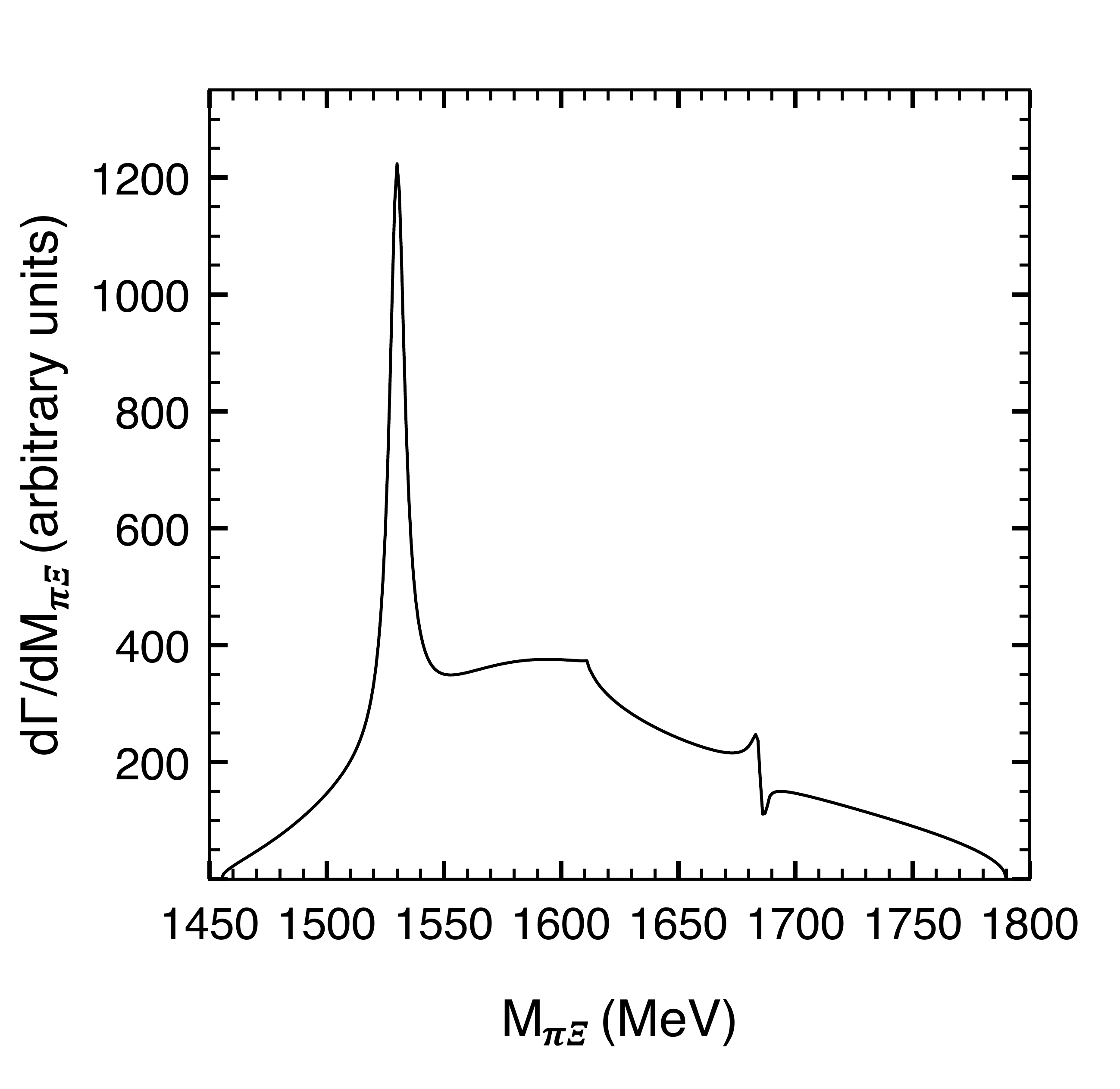}
\caption{The $\pi \Xi$ mass spectrum in the process $\Lambda_c^+ \to  \Xi^- \pi^+ K^+$ obtained by summing a Breit-Wigner distribution 
for the $\Xi(1530)$ resonance to the s-wave $t$-matrix obtained in the present work. It is possible to notice the presence of a structure around 1690 MeV and a $K\Lambda$ cusp around 1612 MeV.}\label{piXispect}
\end{figure}

Before proceeding further, it should be mentioned that  another pole  appears in the spin, isospin 1/2 configuration but deep in the complex plane, around 1530 MeV, with full width  around 200 MeV. A  broad structure around this energy region is seen in the $\pi \Xi$ t-matrix (in Fig.~\ref{ampl_Sh}). However,  identifying such broad states in experimental data should be difficult, especially because a narrow $3/2^+$ resonance, $\Xi(1530)$, exists in this energy region. As an example, we show  the  $\pi \Xi$ mass spectrum in Fig.~\ref{piXispect},
for the process studied by the {\it BABAR} Collaboration, $\Lambda_c^+ \to  \Xi^- \pi^+ K^+$, n\"aively obtained as
\begin{align}
\frac{d\Gamma}{dM_{\pi\Xi}} \propto p_{K} ~p_{\pi \Xi} \mid \mathcal{T}_{\pi \Xi}\mid^2.\label{piXiMS}
\end{align}
Here, $p_K$ and $p_{\pi \Xi}$ represent the momentum of the kaon in the $\Xi^- \pi^+ K^+$ system and the center of mass momentum in the $\pi \Xi$ subsystem, respectively. The matrix element,  $\mathcal{T}_{\pi \Xi}$, in Eq.~(\ref{piXiMS}) is 
\begin{eqnarray}\label{tau}
\mid \mathcal{T}_{\pi \Xi}\mid^2 = \mid T_{\textrm s-wave} \mid^2 +\mid T_{\textrm p-wave} \mid^2%\\\nonumber
\end{eqnarray}
%\begin{eqnarray}\label{tau}
%&&\mid \mathcal{T}_{\pi \Xi}\mid^2 = \mid T_{\textrm s-wave} \mid^2 +\\\nonumber
% &&\sum\limits_{m_{s_i}, m_{s_f}} \int d \Omega \mid \langle S_i, m_{s_i}  \mid \mathcal{T}_{\pi \Xi} \mid S_f, m_{s_f} \rangle\mid^2,
%\end{eqnarray}
%where $S_i$ ($S_f$), $m_{s_i}$ ($m_{s_f}$) represent the spin and the third component of the spin of the baryon in the initial (final) state, and
%\begin{eqnarray}\nonumber
%&& \langle S_i, m_{s_i}  \mid \mathcal{T}_{\pi \Xi} \mid S_f, m_{s_f} \rangle\\\nonumber &&=   \sum\limits_{m_{L_i}, m_{L_f}}\langle S_i, m_{s_i}, L, m_{L_i} \mid J, m \rangle 
%\langle J, m \mid \mathcal{T}_{\pi \Xi} \mid J, m \rangle 
%\\\nonumber &&
%\quad \times \langle J,m \mid S_f, m_{s_f}, L, m_{L_f} \rangle Y_{L, m_{L_i}}Y^*_{L, m_{L_f}}.
%\end{eqnarray}
We calculate Eq.~(\ref{tau})  by using our amplitude for the s-wave part and by  
 writing a  Breit-Wigner function for the p-wave  ($\Xi(1530)$) 
\begin{align}
T_{\textrm p-wave}=\frac{g^2}{\sqrt{s} - M_R + i\Gamma/2}.
\end{align}
We use $M_R = 1530$ MeV and $\Gamma/2 = 5$ MeV for $\Xi(1530)$, following  Ref.~\cite{pdg}, and estimate the coupling $g^2 \sim 0.47$ by assuming the decay rate $\Xi(1530)\to \pi \Xi$ to be nearly 100$\%$ \cite{pdg}. 

\subsection{Isospin = 1/2, Spin =3/2 and Parity = $-1$}
\begin{figure}[h!]
\includegraphics[width=0.5\textwidth]{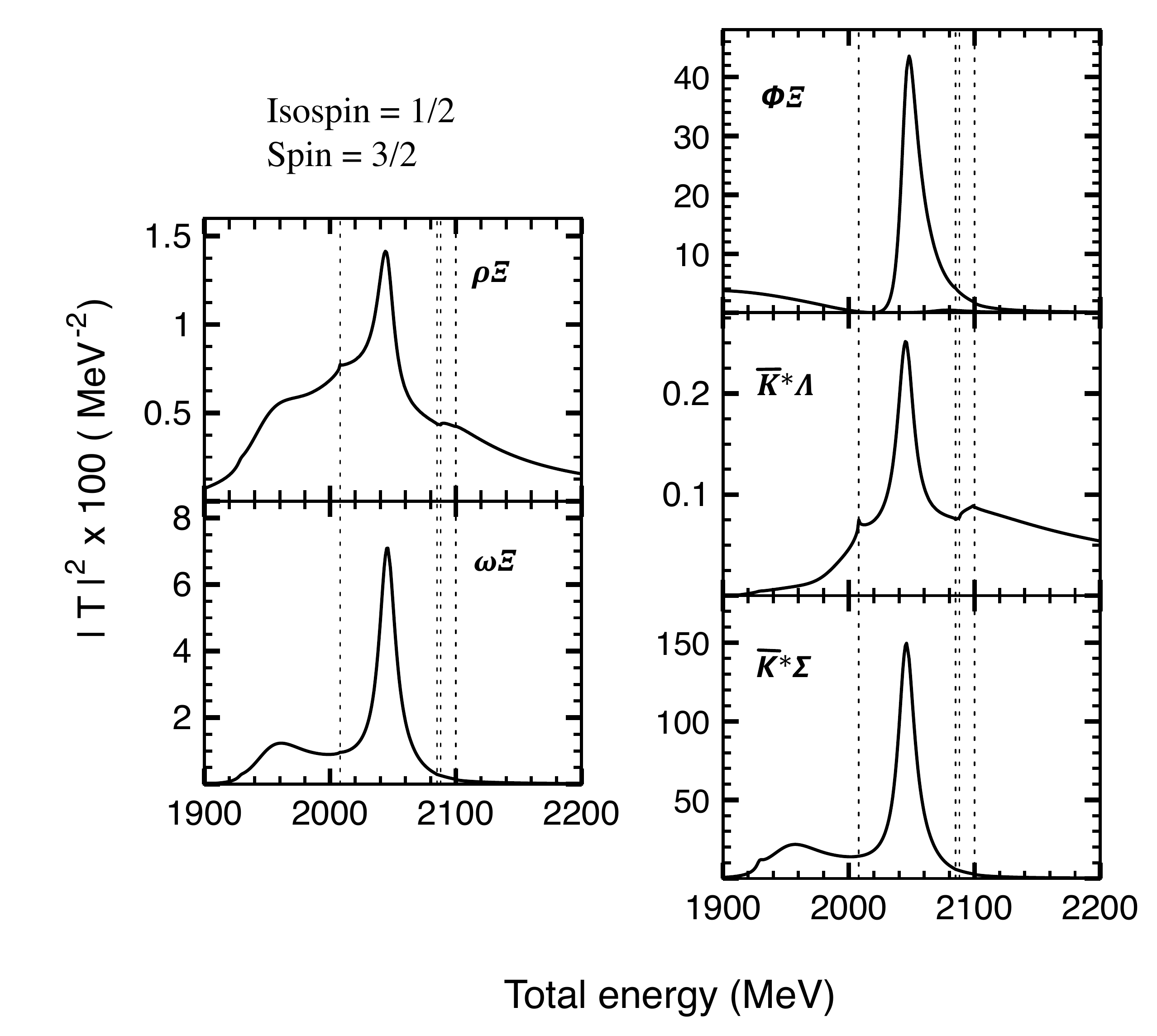}
\caption{Squared amplitudes for total isospin 1/2 and spin 3/2. The vertical dashed lines indicate the thresholds as summarized in Table~\ref{masses}.  }\label{ampl_S3h}
\end{figure}

\begin{figure}[h!]
\includegraphics[width=0.45\textwidth]{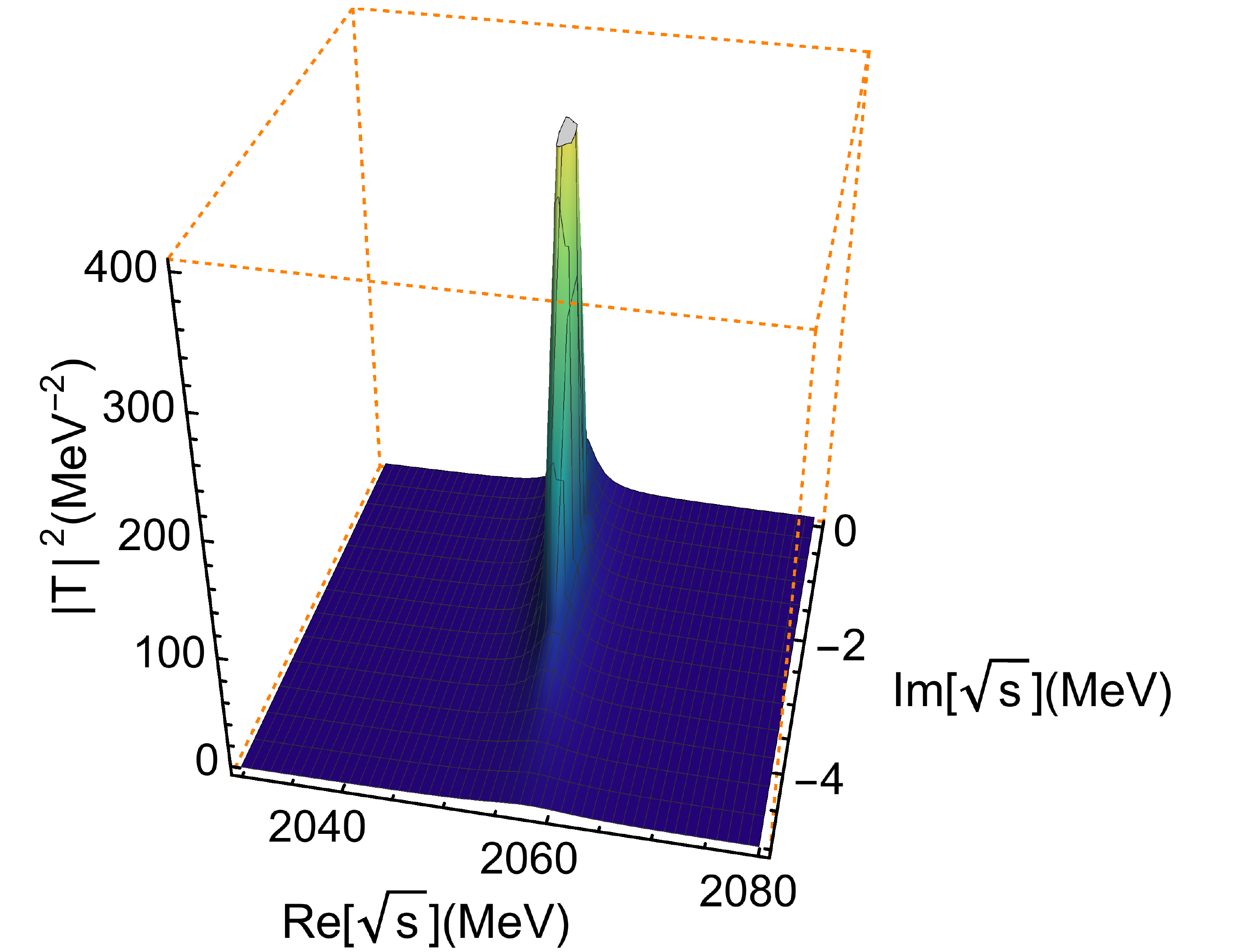}
\caption{Squared amplitude of the $\bar K^* \Sigma$ channel, in the complex plane, calculated by neglecting the widths of the vector mesons.}\label{Xi2050pole}
\end{figure}

Next, we show the squared amplitudes for the isospin 1/2 and spin 3/2 configurations in Fig.~\ref{ampl_S3h}. 
In this case the presence of a  resonance can be clearly seen  around 2050 MeV, which couples strongly to the $\bar K^* \Sigma$ channel. It is important to mention here that a difficulty arises when investigating the pole corresponding to this state in the complex plane, since the widths of unstable mesons like $\rho$, $K^*$ transform a fixed threshold into a band of energies related to a variable  threshold. An appropriate way to tackle this problem would be to calculate the loop functions by including the self-energy of the unstable mesons in the system (for more discussions see, for example, Refs.~\cite{selfenergy}).  Alternatively, an estimation of the effect of the widths of the unstable mesons in the system can be made following the procedure  suggested in Ref.~\cite{ramosvb}, which consists of calculating the mass, width and couplings of the state from the squared amplitudes obtained on the real axis.  We follow this latter procedure and  determine  the mass and half width  of the spin 3/2 state found in our work as 
\begin{equation}\label{Mxi2120} M - i\Gamma/2 = 2046.0 - i 8.2~{\rm MeV.} \end{equation}
The presence of the pole in the complex plane is confirmed by neglecting the widths of  $\rho$ and $K^*$ in the calculations. In such a limit, we do find a pole (as shown in Fig.~\ref{Xi2050pole}) at  
\begin{equation}\label{Pxi2120} M - i\Gamma/2 = 2055 - i 2~{\rm MeV,} \end{equation}  
where the mass value is similar to the one in Eq.~(\ref{Mxi2120}) obtained from the amplitudes shown in Fig.~\ref{ampl_S3h}. 
However, the width in Eq.~(\ref{Pxi2120}) is not well determined since $\rho$ and $K^*$ have been considered as stable particles for the sake of calculations in the complex plane. In any case, Fig.~\ref{Xi2050pole} shows that there exists a pole in the complex plane associated to the peaks seen in squared amplitude in Fig.~\ref{ampl_S3h}. We can, thus, conclude that a $\Xi$ baryon with spin-parity $3/2^-$ arises at  $2046.0 - i 8.2$ MeV due to the vector meson-baryon coupled channel dynamics. Once again we come across a $\Xi$ with a narrow width in spite of the availability of a large phase space to decay. In fact the decay width of this $\Xi$ is mostly through the decay of $\rho$ and $\bar K^*$, and the corresponding peaks in the squared amplitudes tend to get narrower (corresponding to a nearly bound state seen in Fig.~\ref{Xi2050pole}) if the widths of  $\rho$ and $\bar K^*$ is considered to be  zero.

\begin{figure}[h!]
\includegraphics[width=0.22\textwidth]{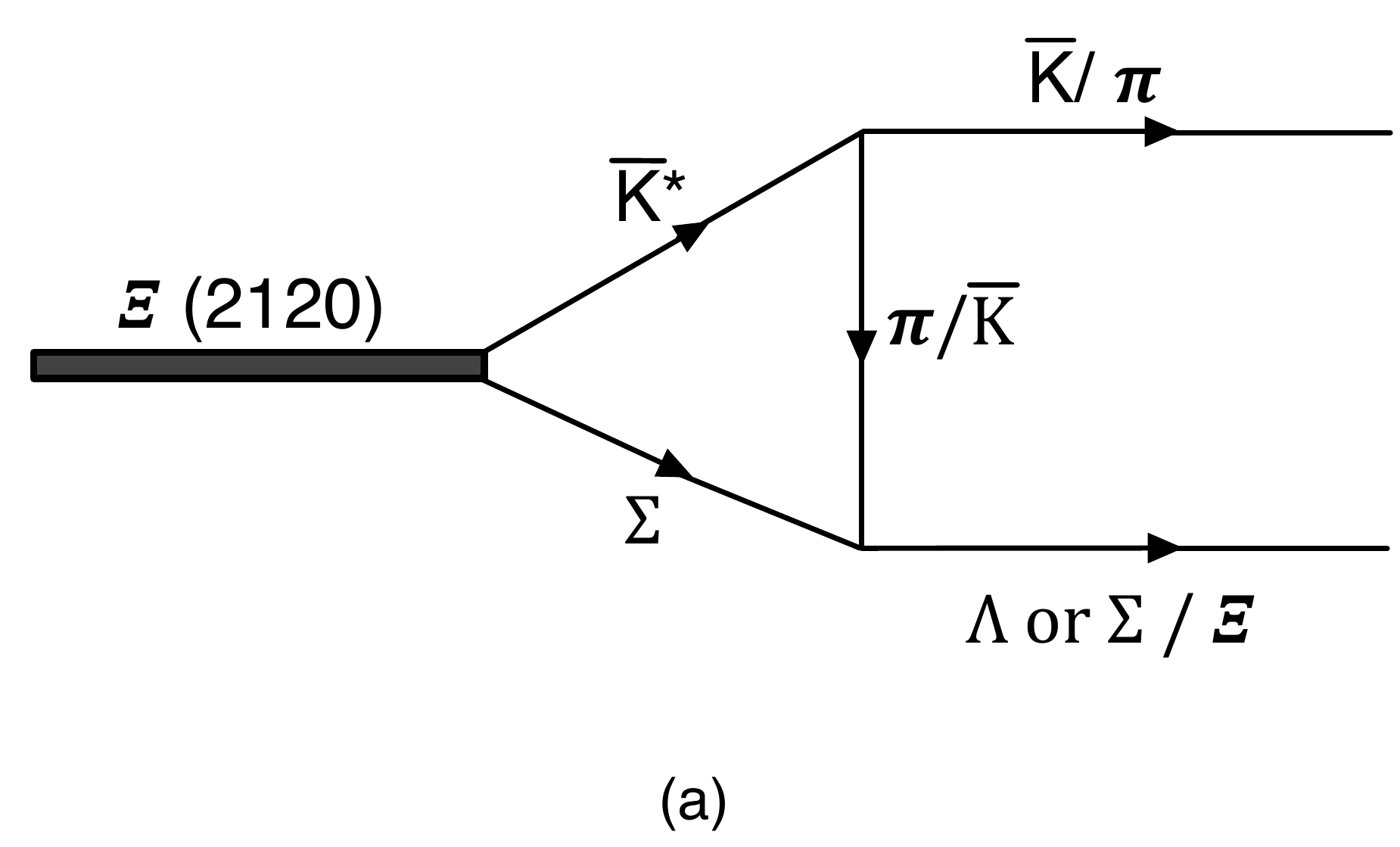}~~~~\includegraphics[width=0.22\textwidth]{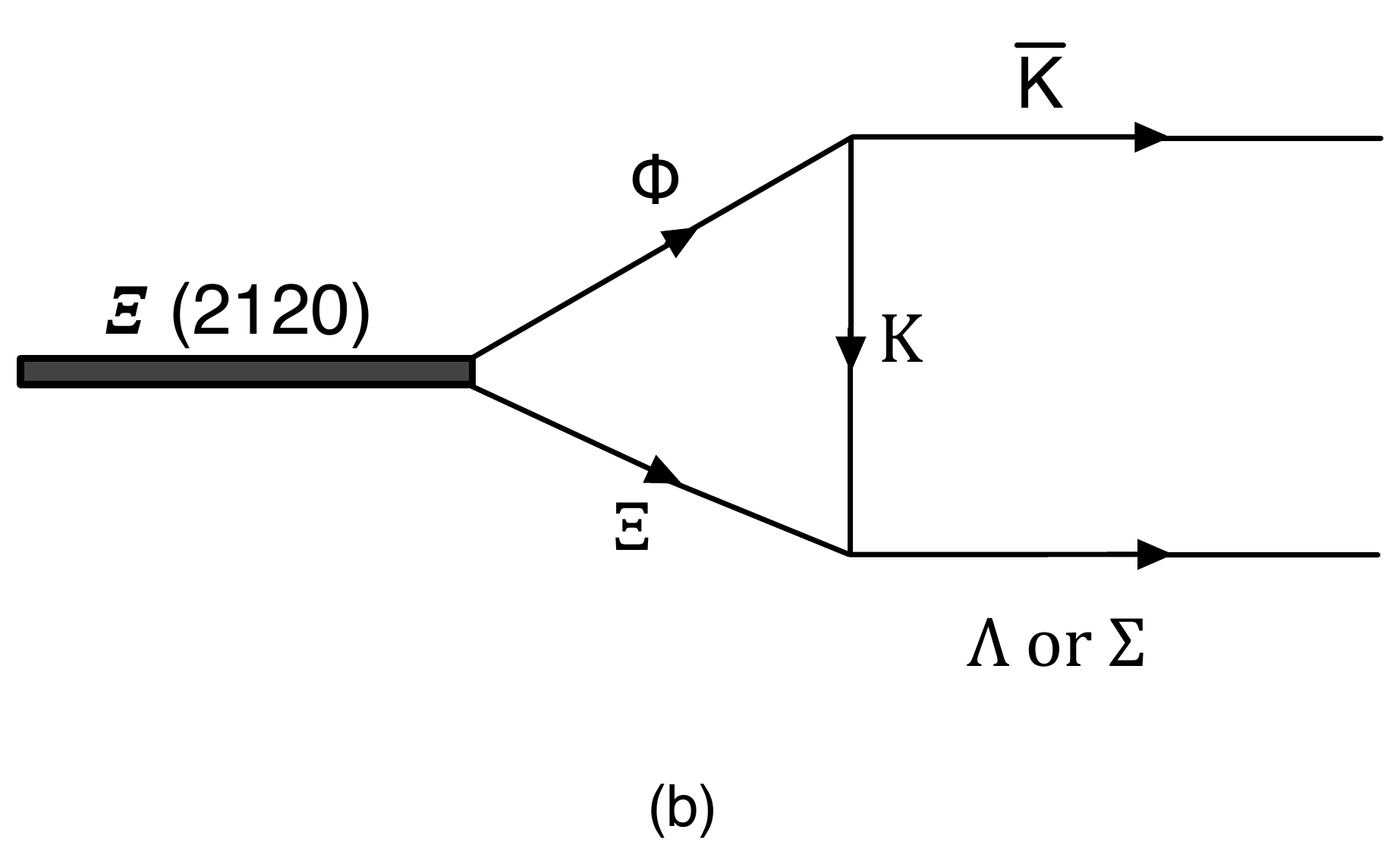}
\caption{Possible decay mechanism of  a $\Xi$ with mass around 2050 MeV to the $\bar K \Lambda$ channel.}\label{decay3}
\end{figure}

The question that arises  is if we can relate this peak structure with any known state. Two poorly known states around 2 GeV are listed in Ref~\cite{pdg}: $\Xi(2030)$ and $\Xi(2120)$. The former one seems to have a spin $J \geq 5/2$~\cite{pdg}. The existence of the latter one is based on a signal found in the $\bar K \Lambda$ mass spectrum~\cite{pdg}.  
Although the mass of the state in Eq.~(\ref{Mxi2120}) is far from that of $\Xi(2120)$, it may be possible to relate the two of them. One reason being the narrow width of $\Xi(2120)$, which is listed as $< 20 $ MeV in Ref.~\cite{pdg}, which is in good agreement with Eq.~(\ref{Mxi2120}). Another reason may be that the $\bar K \Lambda$ channel may not be the most suited channel to investigate the properties of $\Xi(2120)$, if the latter is related to the state found in our work.  As we shall argue now, in such a case,  $\pi \Xi$, $\bar K \Sigma$ channels should be more suitable to study $\Xi(2120)$.  The couplings of the state in Eq.~(\ref{Mxi2120}) to different  channels are given in Table~\ref{coupXi21203/2}. These couplings have been obtained from the amplitudes calculated on the real axis (following Ref.~\cite{ramosvb}). It can be seen from Table~\ref{coupXi21203/2} that the largest coupling of the state are found to  $\bar K^* \Sigma$ and $\phi \Xi$ channels. Under the assumption that our state is related to $\Xi(2120)$, we can say that $\Xi(2120)$ should mainly decay through  $\bar K^* \Sigma$ and $\phi \Xi$ to  final states: $\pi \Xi$, $\bar K \Sigma$, $\bar K \Lambda$, as shown in Fig.~\ref{decay3}.
 It can be anticipated that the amplitudes with final states $\bar K \Sigma$ and $\phi \Xi$ are larger than that for $\bar K \Lambda$ decay by comparing the couplings of the vertices: $\pi \Sigma \Lambda$, $\pi \Sigma \Sigma$, $\bar K \Sigma \Xi$, $K \Xi \Lambda$ and $K \Xi \Sigma$ given in Ref.~\cite{Borasoy}. We list them here for the reader's convenience: the meson-baryon-baryon couplings needed for the $\bar K \Lambda$ final state are $g_{\pi \Sigma \Lambda} =2D$ and  $g_{K \Xi \Lambda} = -D+3F$, in case of $\bar K \Sigma$ channel $g_{\pi \Sigma \Sigma} = 2\sqrt{3}F$ and $g_{K \Xi \Sigma} = \sqrt{3}(D+F)$ contribute and $g_{\bar K \Sigma \Xi} = \sqrt{6}(D+F)$ contributes to the $\pi \Xi$ final state, where $D = 0.8$,  $F=0.46$ \cite{Borasoy}. A more detailed analysis should be made in future but this preliminary comparison suggests that  a $\Xi$ resonance with mass around 2050 MeV and spin-parity $3/2^-$ should be more ideally looked for in the $\pi \Xi$ and $\bar K \Sigma$ mass spectrum.    

\begin{table}[h!]
\caption[]{ Couplings of different channels to $\Xi(2120)$.  
} \label{coupXi21203/2}
\begin{ruledtabular}
\begin{tabular}{cc}
Channels &Couplings   \\
\hline
$\rho \Xi$           &$-0.7 + i 0.0$ \\
$\omega \Xi$&$~1.4 + i 0.0$ \\
$\phi \Xi$          &$-1.9 + i 0.0$ \\
$\bar K^* \Sigma$       &$~3.2 + i 0.0$ \\
$\bar K^*\Lambda$        &$-0.5 + i 0.0$ \\
\end{tabular}
\end{ruledtabular}
\end{table}

We also calculate isospin $3/2$ amplitudes for spin 1/2 and 3/2 but find no resonance in such configurations, thus, implying absence of any exotic states in our study. A claim of the existence of a narrow, exotic $\Xi^{--}$ state, with mass $\sim$ 1862 MeV, in the $\pi^- \Xi^-$ mass spectrum was reported in Ref.~\cite{Alt:2003vb}. However, later investigations have failed to confirm the existence of this state (see Ref.~\cite{Abelev:2014qqa} for some recent results on this topic).

\begin{figure}[h!]
\includegraphics[width=0.4\textwidth, height=6cm]{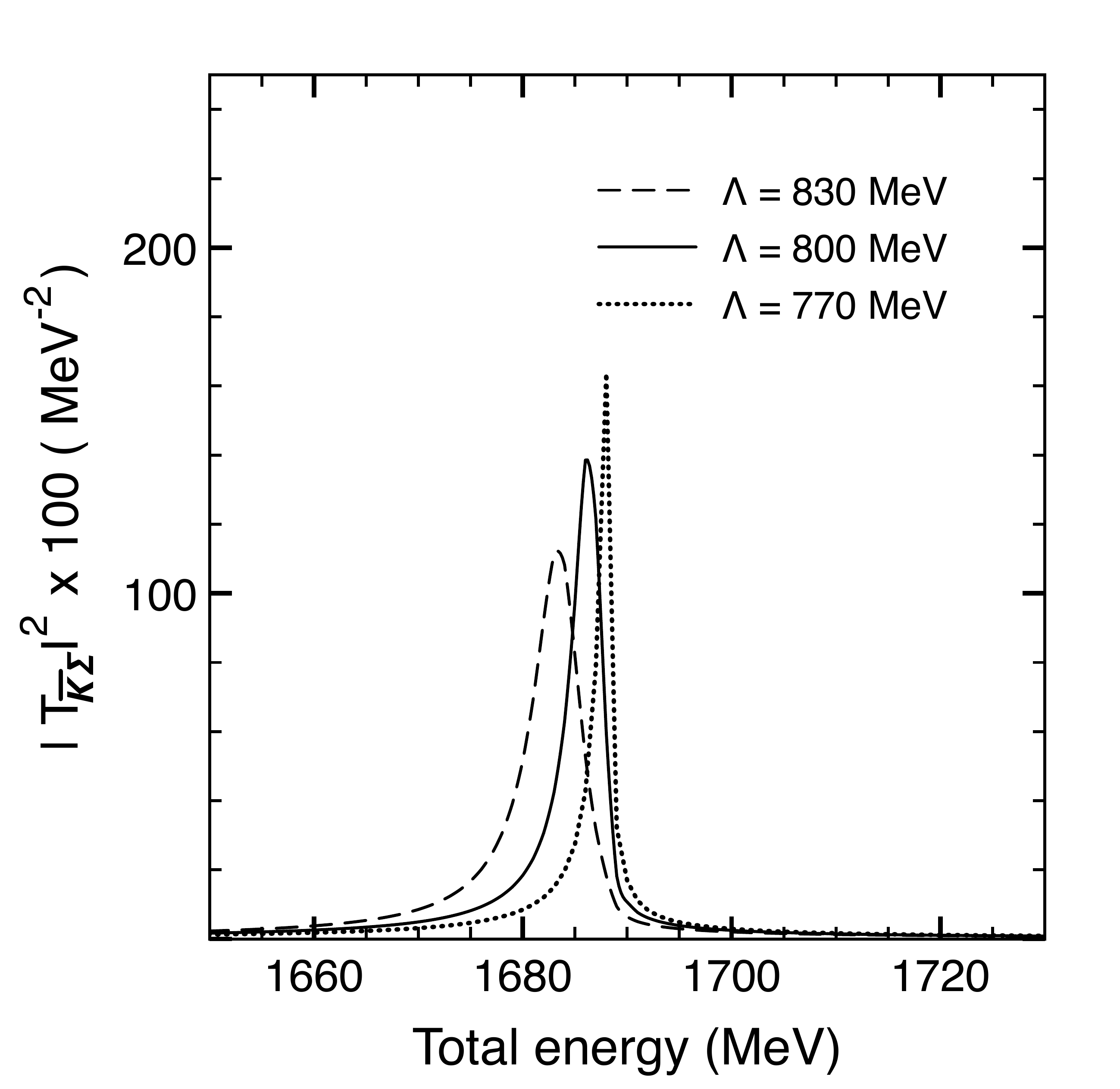}
\caption{Variations in the isospin-spin 1/2 squared amplitude of $\bar K \Sigma$ channel as a function of the cutoff $\Lambda$ used in Eq.~(\ref{loopcutoff}). The peak is associated to $\Xi(1690)$. }\label{Xi1690}
\end{figure}

\begin{figure}[h!]
\includegraphics[width=0.4\textwidth, height=6cm]{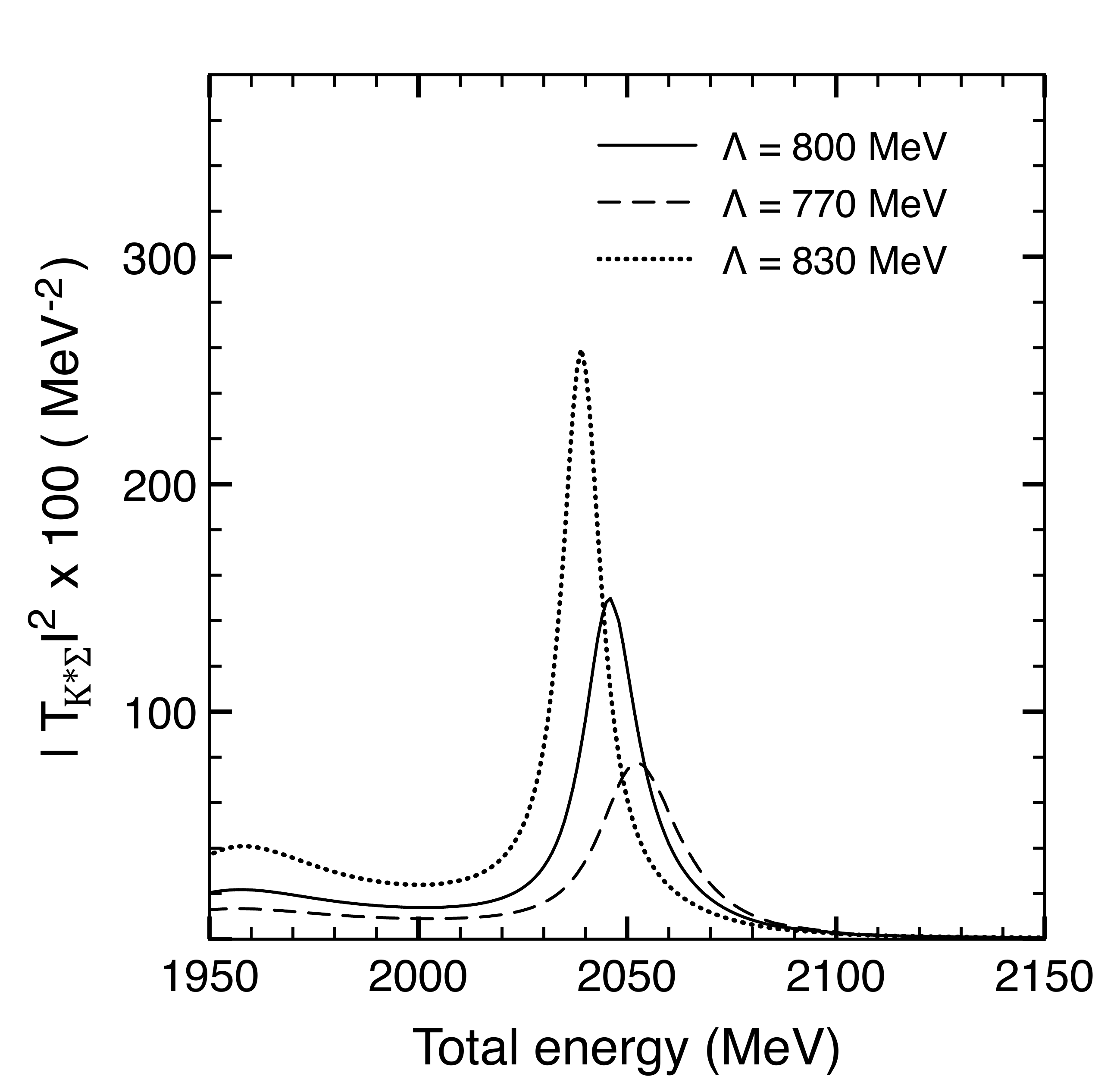}
\caption{Same as Fig.~\ref{Xi1690} but for $\bar K^* \Sigma$ in the isospin 1/2, spin 3/2 configuration. The peak seen in this plot is related to $\Xi(2120)$. }\label{Xi2120}
\end{figure}

Finally, we have checked the stability of our results against the variation of the unique parameter of the calculation, which is the cutoff  ($\Lambda = 800$ MeV) used to calculate the loop function (Eq.~(\ref{loopcutoff})). We change the cutoff $\Lambda$ by $\pm ~4\%$ and  show, as an example, the variation of the squared amplitudes of the $\bar K \Sigma$ and $\bar K^* \Sigma$ channels in Figs~\ref{Xi1690} and \ref{Xi2120}, respectively. 
%\begin{figure}[h!]
%\includegraphics[width=0.4\textwidth, height=6cm]{Xi2120_rho.pdf}
%\caption{Variations in the isospin-spin 1/2 squared amplitude of $\bar K \Sigma$ channel as a function of the cutoff $\Lambda$ used in Eq.~(\ref{loopcutoff}). The peak is associated to $\Xi(1690)$. }\label{Xi2120rho}
%\end{figure}
 The variation in the pole position and half widths of the resonance associated to $\Xi(1690)$ as well as  the $3/2^-$ state related to $\Xi(2120)$ can be summarized as $(1689 \pm 1) + i (2 \pm 1)  $ MeV and $(2046 \pm 6) + i (8.2 \pm 2.2)$ MeV, respectively. It can be seen from these summarized  results that the states found in our work are reasonably stable. %The pole positions in the complex plane for both these states are also stable and show a variation of the order of $\sim \pm 5$ MeV. 
%However, the pole found in the complex plane for the $1/2^-$ state  (Eq.~(\ref{2080}))  related to $\Xi(2120)$ is found to be sensitive to the cutoff. It becomes a threshold effect for $\Lambda = 750$ MeV.

\begin{figure}[h!]
\includegraphics[width=0.4\textwidth]{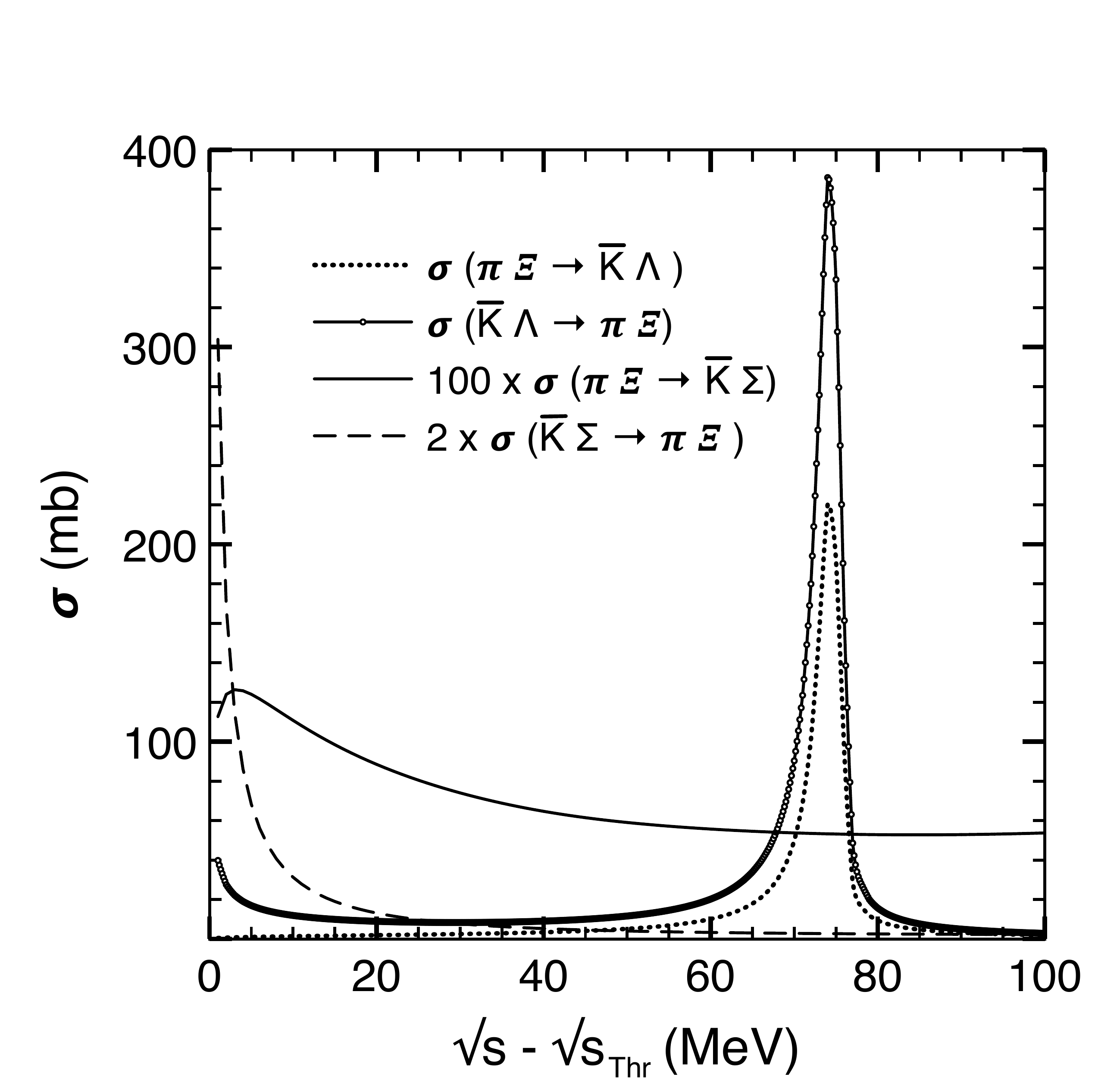}
\caption{Cross sections for the $\pi \Xi \leftrightarrow \bar K \Lambda$, $\pi \Xi  \leftrightarrow \bar K \Sigma$ processes. The horizontal axis corresponds to the total energy above the respective threshold of each process.  We have used isospin 1/2, spin 1/2 amplitudes to obtain these cross sections. }\label{Xns}
\end{figure}

The possibility of formation of resonances, like $\Xi(1690)$, due to meson-baryon coupled channel dynamics, can affect relevant cross sections. Such a modification of cross sections, in turn, can  be useful in understanding the subthreshold $\Xi(1314)$ production observed in Ar+Kcl  collisions at 1.76~A GeV (the threshold in nucleon-nucleon collisions is 3.74 GeV) by the HADES Collaboration \cite{hades}. The measured $\Xi^-$ abundance is more than one order of magnitude larger than predicted by the statistical model \cite{hics} and by the relativistic transport model \cite{Chen:2003nm}. An attempt to explain the observed excess of $\Xi^-$ has been presented in Ref.~\cite{Li:2012bga}, where, with the help of an SU(3) effective Lagrangian model, the authors computed the cross section for the $Y Y \to N \Xi$, $\bar K Y \to \pi \Xi$ and $\bar K N \to K \Xi$ reactions. Including these processes in the RVUU transport model the authors obtained the abundance ratio $R_s = \Xi/(\Lambda + \Sigma^0) = 3.38 \times 10^{-3}$, to be compared with the experimental value of $(5.6 \pm 1.2) \times 10^{-3}$. However, the cross sections used for the process $\bar K \Lambda \to \pi \Xi$ in Ref.~\cite{Li:2012bga} does not account for the $\Xi(1690)$ resonance formation. The inclusion of such an information  drastically enhances the $\bar K \Lambda \to \pi \Xi$ cross sections in the lower energy region and changes the line shape of the energy dependence, which should be important. This is shown in Fig.~\ref{Xns}, 
which should be compared with Fig.~2 of Ref.~\cite{Li:2012bga}. We also show the cross sections for $\pi \Xi  \leftrightarrow \bar K \Sigma$ in Fig.~\ref{Xns}. Using the cross sections obtained here could increase the abundance ratio $R_s$ obtained in Ref.~\cite{Li:2012bga}, bringing it closer to the experimental value.

\section{A comparison of the relativistic and nonrelativistic treatment of the amplitudes}
It is a standard approach to obtain the lowest order amplitudes within a nonrelativistic approximation when studying meson-baryon interactions with the idea of searching for dynamical generation of resonances. In the present case, though, we disregard such an approximation, keeping in mind that the thresholds of the different coupled channels differ up to $\sim 200$ MeV. We find it useful to end this article with a comparison of the different amplitudes obtained within a nonrelativisitic approximation with those depicted in Figs.~\ref{ampl_Sh}~and~\ref{ampl_S3h}. To be explicit, we wish to compare the squared $t$-matrices shown in Figs.~\ref{ampl_Sh}~and~\ref{ampl_S3h} with those obtained by solving the Bethe-Salpeter equation with the kernels, $V_{VB}$, deduced in the nonrelativistic limit, while the loop function $G$ in Eq.~(\ref{BS}) continues to be relativistic. Under the nonrelativisitic approximation (considering energy of the baryon $E \sim M$, and neglecting the time component of the polarization vector of the vector meson), the contact term, for example, becomes
\begin{eqnarray}
V^{I,S}_{\textrm{CT,VB}} &=&i \,C^{I}_{\textrm{CT,VB}}\,\frac{g_1 g_2}{2 \sqrt{M_1 M_2}} \,\,\vec{\sigma}\cdot\vec{\epsilon_2} \times \vec{\epsilon_1},\label{VctNR}
\end{eqnarray}
where the meaning of  $C^{I}_{\textrm{CT,VB}}$, $g$ and $M_1$, $M_2$ is the same as in Eq.~(\ref{contacts}). Consistently, neglecting the terms of order $\mathcal O (|\vec p|/{\rm mass})$ (modulus of momentum divided by mass), or bigger, the $t$-channel amplitude given by Eq.~(\ref{Vtex}) reduces to
\begin{eqnarray}
%V^{I}_{t, {\rm VB}} &=&  \frac{C_{t, {\rm VB}}^{I}}{4 f_V^2} \left(2 \sqrt{s} - M_1 - M_2 \right) \\
%&\times&\sqrt{\frac{M_1 + E_1}{2 M_1}} \sqrt{\frac{M_2 + E_2}{2 M_2}}\,\,\epsilon_1\cdot \epsilon_2.%-\frac{C^I_{t, VB}}{4 f_{V}^2}(\omega + \omega^\prime)  \vec{\epsilon}_1\cdot \vec{\epsilon}_2,
V^{I}_{t, {\rm VB}} &=& \frac{-m_{Vx}^2}{4 f_{Vi} f_{Vj}}\frac{1}{t-m_{Vx}^2}\Biggl\{ \epsilon_1\cdot\epsilon_2 \biggl[ \biggl( 2 \sqrt{s} - M_1 - M_2 \biggr.\biggr.\Biggr. \\\nonumber
&+&  \left. (M_1-M_2)\frac{(m_2^2-m_1^2)}{m_{Vx}^2} \right) C_{t1, {\rm VB}}^{I} \\\nonumber
&+& \biggl. \Biggl. \left( \frac{M_1+ M_2}{2 M}\left(2 \sqrt{s} - M_1 - M_2\right)-\frac{s-u}{2M}\right)C_{t2, {\rm VB}}^{I}
\biggr] \Biggr\}, 
\end{eqnarray}
which can be further simplified by considering that: (1) the term multiplied by $C_{t2, {\rm VB}}^{I}$ is much smaller than the one multiplied by $C_{t1, {\rm VB}}^{I}$ (approximating $M_1 \simeq M_2 \simeq M$, $m_1 \simeq m_2 \simeq m$, implying $s - u \sim 4 M m$) and can be neglected; (2) the term with $(m_2^2-m_1^2)/m_{Vx}^2$ is negligible as compared to the remaining terms; (3) $t - m_{Vx}^2 \to - m_{Vx}^2$. These approximations simplify the $t$-channel amplitude to
\begin{eqnarray}
V^{I}_{t, {\rm VB}} &=& \frac{-\vec \epsilon_1\cdot\vec \epsilon_2}{4 f_{Vi} f_{Vj}}  \biggl[ \left( 2 \sqrt{s} - M_1 - M_2 \right) C_{t1, {\rm VB}}^{I}\biggr].\label{VtNR}
\end{eqnarray}
The amplitude in Eq.~(\ref{VtNR}) corresponds to the one used in Refs.~\cite{Kolomeitsev:2003kt,GarciaRecio:2003ks,ramosvb,Gamermann,Pavon,vbvb,pbvb,hyperons}
\begin{eqnarray}
V^{I}_{t, {\rm VB}} =-\frac{C^I_{t1, VB}}{4 f_{Vi} f_{Vj}} (\omega + \omega^\prime)  \vec{\epsilon}_1\cdot \vec{\epsilon}_2,
\end{eqnarray}
where $\omega + \omega^\prime$ represents the sum of the energies of the incoming and outgoing meson, which, in the center of mass, is
$ \omega + \omega^\prime = 2 \sqrt{s} - E_1- E_2 \simeq  2 \sqrt{s} - M_1- M_2$ in the nonrelativistic approximation. Analogously, one can easily reduce the lowest order amplitudes for the $s$- and $u$-channel to 
\begin{eqnarray}
%V^I_u&=-C^I_u\left(\frac{g^2}{2\bar M-m}\right) ,\nonumber\\
V^{I,S}_{s, {\textrm VB}}&=&C^I_{s, {\textrm VB}} \left(\frac{g^2}{2\bar M + \bar m}\right) \vec{\epsilon}_2\cdot \vec{\sigma}\,\, \vec{\epsilon}_1\cdot \vec{\sigma },\label{VsNR}\\
%V^I_\textrm{CT}&=C^I_{\textrm{CT}}\,\frac{g^2}{M},\label{Vuc1}
V^{I,S}_{u, {\textrm VB}}&=& C^I_{u, {\textrm VB}} \left(\frac{g^2}{2\bar M - \bar m}\right) \vec{\epsilon}_1\cdot \vec{\sigma}\,\, \vec{\epsilon}_2\cdot \vec{\sigma },\label{VuNR}
\end{eqnarray}
where $\bar M$  and $\bar m$ are SU(3) average masses of the baryon and vector meson and 
\begin{eqnarray}\nonumber
 C^I_{s, {\textrm VB}} &=& I^s_{1f} I^s_{1i} - \frac{\bar m}{2 \bar M}\left( I^s_{1f} I^s_{2i} + I^s_{2f} I^s_{1i} \right) + \left(\frac{\bar m}{2 \bar M}\right)^2 I^s_{2f} I^s_{2i},\\\nonumber
 C^I_{u, {\textrm VB}} &=& I^u_{1f} I^u_{1i} + \frac{\bar m}{2 \bar M}\left( I^u_{1f} I^u_{2i} + I^u_{2f} I^u_{1i} \right) + \left(\frac{\bar m}{2 \bar M}\right)^2 I^u_{2f} I^u_{2i}.
\end{eqnarray}
\begin{figure}[h!]
\includegraphics[width=0.51\textwidth]{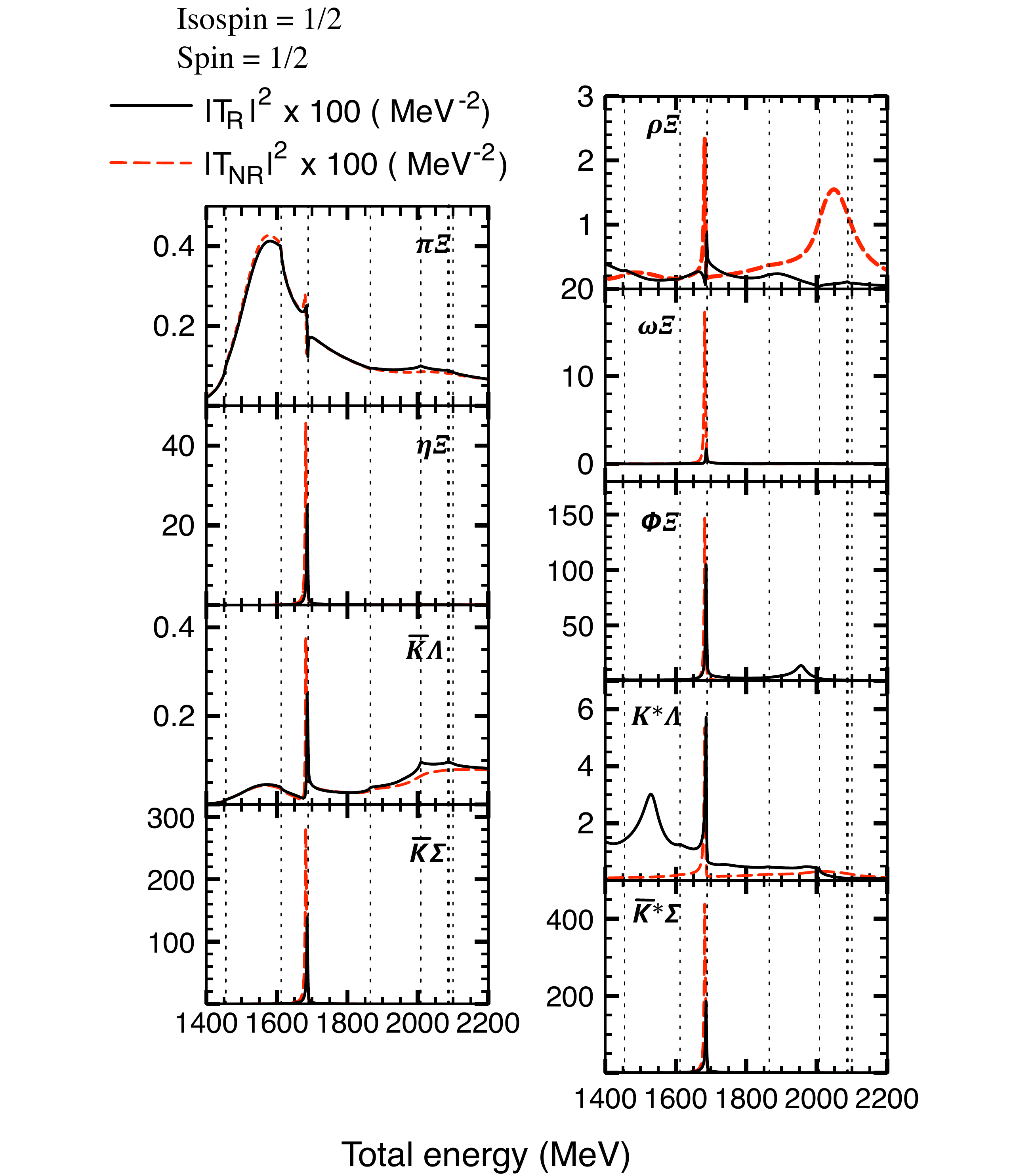}
\caption{Squared amplitudes for the total isospin and spin 1/2. The solid (dashed) curves show the results of the calculations done with  relativistic (nonrelativistic) lowest order amplitudes obtained with cutoff $\Lambda = 800$ MeV. The vertical dashed lines indicate the thresholds  summarized in Table~\ref{masses}. }\label{fig:compare1}
\end{figure}

It can be seen from the comparison of the isospin and spin 1/2 amplitudes in Fig.~\ref{fig:compare1} that the presence of the pole and its position corresponding to $\Xi(1690)$  does not get affected by the nonrelativistic approximation, although the magnitude of the squared amplitudes is different within such an approximation.  Yet another difference is the presence of another bump like structure near 2050 MeV in the $\rho \Xi$ amplitude which does not correspond to a pole in the complex plane but a small variation in the cutoff value can produce a pole corresponding to this peak. We can, thus, summarize that a state with isospin, spin 1/2 and mass near 2050 MeV is not stable and depends on a relativistic or nonrelativistic treatment of the lowest order amplitudes. 

\begin{figure}[h!]
\includegraphics[width=0.51\textwidth]{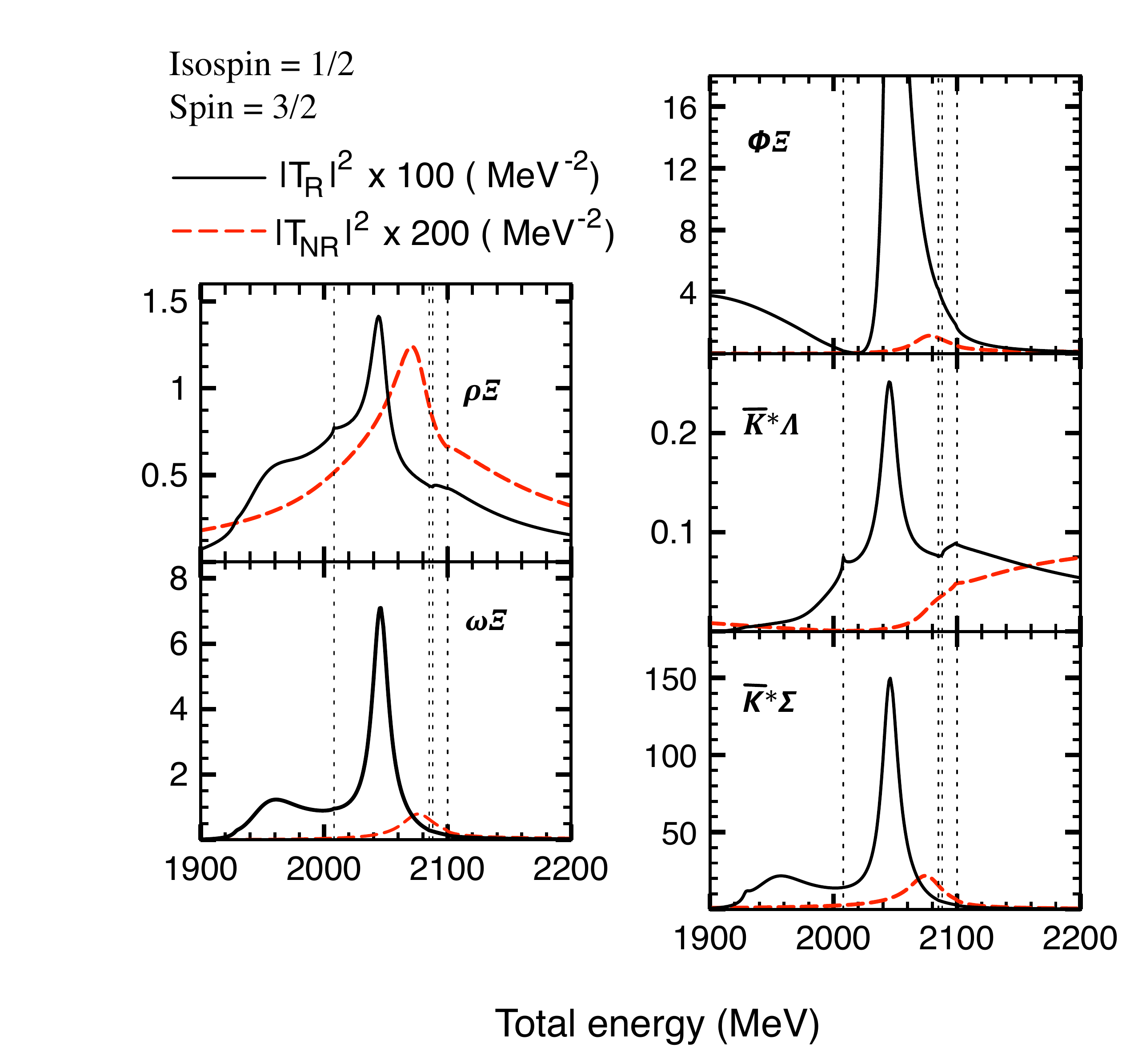}
\caption{Squared amplitudes for the total isospin 1/2 and spin 3/2. The solid (dashed) curves show the results of the calculations done with  relativistic (nonrelativistic) lowest order amplitudes obtained with cutoff $\Lambda = 800$ MeV. Notice that the dashed curves have been multiplied by an extra factor 2 for the sake of comparison. }\label{fig:compare2}
\end{figure}

We show the comparison of the results obtained within a nonrelativistic approximation considered for the lowest order amplitudes  for the meson-baryon system in isospin 1/2 and spin 3/2 configuration in Fig.~\ref{fig:compare2}. In this case, we can say that the presence of the pole is not affected by the nonrelativistic approximation but the pole positions differs by about 30 MeV and its coupling to the $\bar K^* \Lambda$ channel gets too weak to show  any signal of the resonance.
Like the isospin, spin 1/2 case, the strength of the squared amplitude gets weaker within a nonrelativisitic approximation.  We can summarize this last discussion by mentioning that the nonrelativistic approximation is reliable as far as the nature of a dynamically generated state is concerned. However, the strength of the amplitudes and pole positions can get affected  by this latter approximation.

\section{Summary}
We can summarize this manuscript by mentioning that we have shown that pseudoscalar-baryon and vector-baryon coupled channel dynamics generate a resonance which can be well associated to $\Xi(1690)$. The present formalism can explain all the recent experimental findings related to $\Xi(1690)$ with a single parameter of the calculations; a cutoff required to regularize the loop functions. We also find a narrow  $3/2^-$ state with mass $\sim$ 2050 MeV, and relate it to  $\Xi(2120)$. We show that the narrow widths of $\Xi(1690)$ and $\Xi(2120)$ can be understood in terms of their weak couplings to the open channels, which is a result of intricate coupled channel meson-baryon dynamics. It is important to mention here that the vector-baryon dynamics play an essential role in the generation of these poles and in explaining the relevant experimental data. We find that the $t$-channel interaction and the contact term give dominant contributions to the VB amplitudes.
Finally, the existence of such cascades in nature might be an interesting information in statistical models trying to understand the ratio of  $\Xi^-/\Lambda$ production found in heavy ion collisions \cite{hades}. We show the cross sections for the processes $\pi \Xi \leftrightarrow \bar K \Lambda$, $\pi \Xi  \leftrightarrow \bar K \Sigma$  
which can  be useful in studying corresponding in-medium processes. 

\section*{Acknowledgements}
K.P.K, A.M.T, F.S.N and M.N gratefully acknowledge the financial support received from FAPESP (under  the
grant number 2012/50984-4) and CNPq (under the grant numbers 310759/2016-1 and 311524/2016-8). A.H and H.N thank the support from Grants-in-Aid for Scientific Research (Grants No. JP17K05441(C)) and  (Grants No. JP17K05443(C)), respectively.

\appendix
\setcounter{table}{0}
\renewcommand{\thetable}{A\arabic{table}}
\section{Isospin coefficients for different amplitudes}\label{coeff}
\begin{table} [h!]
\caption{ $C^{1/2}_{\textrm{CT,VB}}$  coefficients of the contact term.} \label{vbvbCT1/2}
\begin{ruledtabular}
\begin{tabular}{cccccc}
&$\rho \Xi$&$\omega \Xi$&$\phi \Xi$&$\bar K^* \Sigma$&$\bar K^* \Lambda$\\
\hline
$\rho \Xi$&$\left( F - D \right)$&$0$&$0$&$- \frac{1}{4}\left( D + F \right) $&$-\frac{1}{4}\left( D - 3F \right)$\\
$\omega \Xi$&&$0$&$0$&$- \frac{\sqrt{3}}{4} \left( D + 3F \right) $&$\frac{1}{4 \sqrt{3}}\left( D - 3F \right) $\\
$\phi \Xi$&&&$0$&$-\sqrt{\frac{3}{2}} \frac{\left( D + F \right)}{2} $&$-\frac{1}{2\sqrt{6}}\left( D - 3F \right) $\\
$\bar K^* \Sigma$&&&&$\frac{1}{2}\left( D + 2F \right)$&$-\frac{D}{2}$\\
$\bar K^*  \Lambda$& & &&&$-\frac{D}{2}$\\
\end{tabular}
\end{ruledtabular}
\end{table}
\begin{table} [h!]
\caption{ $C^{3/2}_{\textrm{CT,VB}}$  coefficients of the contact term.} \label{vbvbCT3/2}
\begin{ruledtabular}
\begin{tabular}{ccc}
&$\rho \Xi$&$\bar K^* \Sigma$\\
\hline
$\rho \Xi$&$\frac{\left( D - F \right)}{2}$&$-\frac{\left( D + F \right)}{2}$\\
$\bar K^*\Sigma$& &$\frac{\left( D - F \right)}{2}$\\
\end{tabular}
\end{ruledtabular}
\end{table}
\begin{table} [h!]
\caption{ The product $I^s_{1f}I^s_{1i}$ of the isospin coefficients of the $s$-channel amplitude given by Eq.~(\ref{schannel3}).} \label{i1fi1iS}
\begin{ruledtabular}
\begin{tabular}{cccccc}
&$\rho \Xi$&$\omega \Xi$&$\phi \Xi$&$\bar K^* \Sigma$&$\bar K^* \Lambda$\\
\hline
$\rho \Xi$&$\frac{3}{4}$&$-\frac{\sqrt{3}}{4}$&$ -\sqrt{\frac{3}{2}}$&$\frac{3}{4}$&$-\frac{3}{4}$\\
$\omega \Xi$&&$\frac{1}{4}$&$\frac{1}{\sqrt{2}}$&$-\frac{\sqrt{3}}{4}$&$\frac{\sqrt{3}}{4}$ \\
$\phi \Xi$&&& 2 &$-\sqrt{\frac{3}{2}}$&$\sqrt{\frac{3}{2}}$ \\
$\bar K^* \Sigma$&&&&$\frac{3}{4}$&$-\frac{3}{4}$\\
$\bar K^*  \Lambda$& & &&&$\frac{3}{4}$\\
\end{tabular}
\end{ruledtabular}
\end{table}
\begin{table} [h!]
\caption{The product $I^s_{1f}I^s_{2i}$ of the isospin  coefficients of Eq.~(\ref{schannel3}).} \label{i1fi2iS}
\begin{ruledtabular}
\begin{tabular}{cccccc}
&$\rho \Xi$&$\omega \Xi$&$\phi \Xi$&$\bar K^* \Sigma$&$\bar K^* \Lambda$\\
\hline
$\rho \Xi$&$\frac{3(F-D)}{4}$&$\frac{\sqrt{3}(D - F)}{4} $&$ \sqrt{\frac{3}{2}}(D - F)$&$\frac{3(F - D)}{4}$&$\frac{3(D - F)}{4}$\\
$\omega \Xi$&&$\frac{(F - D)}{4}$&$\frac{(F - D)}{\sqrt{2}}$&$ \frac{\sqrt{3}(D - F)}{4}$&$\frac{\sqrt{3}(F - D)}{4}$ \\
$\phi \Xi$&&&$2F$&$-\sqrt{\frac{3}{2}}F$&$\sqrt{\frac{3}{2}}F$ \\
$\bar K^* \Sigma$&&&&$\frac{3(D+F)}{4}$&$-\frac{3(D+F)}{4}$\\
$\bar K^*  \Lambda$& & &&&$\frac{(3F-D)}{4}$\\
\end{tabular}
\end{ruledtabular}
\end{table}
\begin{table} [h!]
\caption{ The product $I^s_{2f}I^s_{1i}$ of the isospin coefficients of Eq.~(\ref{schannel3}).} \label{i2fi1iS}
\begin{ruledtabular}
\begin{tabular}{cccccc}
&$\rho \Xi$&$\omega \Xi$&$\phi \Xi$&$\bar K^* \Sigma$&$\bar K^* \Lambda$\\
\hline
$\rho \Xi$&$\frac{3(F-D)}{4}$&$\frac{\sqrt{3}(D - F)}{4} $&$ -\sqrt{\frac{3}{2}} F$&$\frac{3(D + F)}{4}$&$\frac{(D - 3F)}{4}$\\
$\omega \Xi$&&$\frac{(F - D)}{4}$&$\frac{F}{\sqrt{2}}$&$ -\frac{\sqrt{3}(D + F)}{4}$&$-\frac{(D - 3F)}{4 \sqrt{3}}$ \\
$\phi \Xi$&&&$2F$&$-\sqrt{\frac{3}{2}}(D+F)$&$-\frac{D- 3F}{\sqrt{6}}$ \\
$\bar K^* \Sigma$&&&&$\frac{3(D+F)}{4}$&$\frac{(D-3F)}{4}$\\
$\bar K^*  \Lambda$& & &&&$\frac{(3F-D)}{4}$\\
\end{tabular}
\end{ruledtabular}
\end{table}
\begingroup
\squeezetable
\begin{table} [h!]
\caption{ The product $I^s_{2f}I^s_{2i}$ of the isospin coefficients of Eq.~(\ref{schannel3}).} \label{i2fi2iS}
\begin{ruledtabular}
\begin{tabular}{cccccc}
&$\rho \Xi$&$\omega \Xi$&$\phi \Xi$&$\bar K^* \Sigma$&$\bar K^* \Lambda$\\
\hline
$\rho \Xi$&$\frac{3(D - F)^2}{4}$&$\frac{-\sqrt{3}(D - F)^2}{4} $&$ \sqrt{\frac{3}{2}} F(D - F)$&$-\frac{3(D^2 - F^2)}{4}$&$\frac{(-D^2 + DF - 3F^2)}{4}$\\
$\omega \Xi$&&$\frac{(D - F)^2}{4}$&$\frac{F (F - D)}{\sqrt{2}}$&$ \frac{\sqrt{3}(D^2 - F^2)}{4}$&$\frac{(D - 3F)(D - F)}{4 \sqrt{3}}$ \\
$\phi \Xi$&&&$2F^2$&$-\sqrt{\frac{3}{2}}F(D + F)$&$-\frac{F(D - 3F)}{\sqrt{6}}$ \\
$\bar K^* \Sigma$&&&&$\frac{3(D+F)^2}{4}$&$\frac{(D^2 - 2DF - 3F^2)}{4}$\\
$\bar K^*  \Lambda$& & &&&$\frac{(D - 3F)^2}{12}$\\
\end{tabular}
\end{ruledtabular}
\end{table}
\endgroup
\begin{table} [h!]
\caption{ $C^{I}_{t1, VB}$  coefficients appearing in the VB $t$-channel amplitude of Eq.~(\ref{Vtex}) for the isospin 1/2.} \label{t1h}
\begin{ruledtabular}
\begin{tabular}{cccccc}
&$\rho \Xi$&$\omega \Xi$&$\phi \Xi$&$\bar K^* \Sigma$&$\bar K^* \Lambda$\\
\hline
$\rho \Xi$      &$2$     &0     &0    &$-\frac{1}{2}$            &$-\frac{3}{2}$\\
$\omega \Xi$ &          &0     &0    &$\frac{\sqrt{3}}{2}$    &$-\frac{\sqrt{3}}{2}$ \\
$\phi \Xi$       &           &      &0    &$-\sqrt{\frac{3}{2}}$  &$\sqrt{\frac{3}{2}}$ \\
$\bar K^* \Sigma$&    &       &     &$2$                           &$0$\\
$\bar K^*  \Lambda$& &      &     &                                 &$0$\\
\end{tabular}
\end{ruledtabular}
\end{table}
\begin{table} [h!]
\caption{ $C^{I}_{t2, VB}$  coefficients appearing in the VB $t$-channel amplitude of Eq.~(\ref{Vtex}) for the isospin 1/2.} \label{t2h}
\begin{ruledtabular}
\begin{tabular}{cccccc}
&$\rho \Xi$&$\omega \Xi$&$\phi \Xi$&$\bar K^* \Sigma$&$\bar K^* \Lambda$\\
\hline
$\rho \Xi$      &$2\left(F-D\right)$     &0     &0    &$-\frac{\left(D+F\right)}{2}$              &$\frac{\left(D - 3F\right)}{2}$\\
$\omega \Xi$ &                                &0     &0    &$\frac{\sqrt{3}\left(D+F\right)}{2}$    &$\frac{\left(D - 3F\right)}{2\sqrt{3}}$ \\
$\phi \Xi$       &                                &       &0    &$-\sqrt{\frac{3}{2}}\left(D+F\right)$  &-$\frac{\left(D-3F\right)}{\sqrt{6}}$ \\
$\bar K^* \Sigma$&                         &       &      &$\left(D+2F\right)$                           &$-D$\\
$\bar K^*  \Lambda$&                     &      &       &                                                        &$-D$\\
\end{tabular}
\end{ruledtabular}
\end{table}
\begin{table} [h!]
\caption{ $C^{I}_{t1, VB}$  coefficients of the VB $\rightarrow$ VB $t$-channel amplitude (Eq.~(\ref{Vtex})) in isospin 3/2.} \label{t13h}
\begin{ruledtabular}
\begin{tabular}{ccc}
&$\rho \Xi$&$\bar K^* \Sigma$\\
\hline
$\rho \Xi$              &$-1$       &$-1$\\
$\bar K^* \Sigma$ &              &$-1$\\
\end{tabular}
\end{ruledtabular}
\end{table}
\begin{table} [h!]
\caption{ $C^{I}_{t2, VB}$  coefficients of the VB $\rightarrow$ VB $t$-channel amplitude (Eq.~(\ref{Vtex})) in isospin 3/2.} \label{t23h}
\begin{ruledtabular}
\begin{tabular}{ccc}
&$\rho \Xi$&$\bar K^* \Sigma$\\
\hline
$\rho \Xi$              &$\left(D - F\right)$       &$-\left(D + F\right)$\\
$\bar K^* \Sigma$ &                                   &$\left(D - F\right)$\\
\end{tabular}
\end{ruledtabular}
\end{table}
\begin{table} [h!]
\caption{ $C^{3/2}_{\rm PBVB}$  coefficients of the PB $\rightarrow$ VB amplitude (Eq.~(\ref{KR})).} \label{kr_iso3/2}
\begin{ruledtabular}
\begin{tabular}{ccc}
&$\rho \Xi$&$\bar K^* \Sigma$\\
\hline
$\pi \Xi$&$\left( F^\prime - D^\prime \right)$&$\left( D^\prime + F^\prime \right)$\\
$\bar K\Sigma$&$\left( D^\prime + F^\prime \right) $ &$\left( F^\prime - D^\prime \right)$\\
\end{tabular}
\end{ruledtabular}
\end{table}
{\small
\begin{table*} [tb]
\caption{ $C^{1/2}_{\rm PBVB}$  coefficients of the PB $\rightarrow$ VB amplitude (Eq.~(\ref{KR})).} \label{kr_iso1/2}
\begin{ruledtabular}
\begin{tabular}{cccccc}
&$\rho \Xi$&$\omega \Xi$&$\phi \Xi$&$\bar K^* \Sigma$&$\bar K^* \Lambda$\\
\hline
$\pi \Xi$&$2\left( D^\prime - F^\prime \right)$&$0$&$0$&$\frac{1}{2}\left( D^\prime + F^\prime \right) $&$-\frac{1}{2}\left( D^\prime - 3F^\prime \right)$\\
$\eta \Xi$&$0$&$0$&$0$&$- \frac{3}{2} \left( D^\prime + 3F^\prime \right) $&$-\frac{1}{2}\left( D^\prime - 3F^\prime \right) $\\
$\bar K\Sigma$&$\frac{1}{2}\left( D^\prime + F^\prime \right) $ &$\frac{-\sqrt{3}}{2} \left( D^\prime + F^\prime \right)$&$\sqrt{\frac{3}{2}} \left( D^\prime + F^\prime \right)$&$-\left( D^\prime + 2F^\prime \right)$&$D^\prime$\\
$\bar K \Lambda$&$-\frac{1}{2} \left( D^\prime - 3F^\prime \right)$ & $-\frac{1}{2\sqrt{3}}\left( D^\prime - 3F^\prime \right)$&$\frac{1}{\sqrt{6}}\left( D^\prime - 3F^\prime \right)$&$D^\prime$&$D^\prime$\\
\end{tabular}
\end{ruledtabular}
\end{table*}}

\begin{minipage}{\linewidth}
Next, we list the definitions of the coefficients $V_{u1},~...V_{u12}$ of Eq.~(\ref{u-channel}):
\end{minipage}
\begin{eqnarray}\label{u1}
V_{u1}&=& \left(I^{u}_{1f} - \frac{ I^{u}_{2f} K_1^0}{2\sqrt{M_1 M_2}}\right)\left(I^{u}_{1i} - \frac{ I^{u}_{2i} K_2^0}{2\sqrt{M_1 M_2}}\right) \\\nonumber
&-& \frac{|\vec{K}_1|^2 I^{u}_{2f}}{4\sqrt{M_1M_2}}\left( I^{u}_{1i} -  \frac{I^{u}_{2i} K_2^0}{2\sqrt{M_1 M_2}}\right) \\\nonumber
&-& \frac{|\vec{K}_2|^2 I^{u}_{2i}}{4\sqrt{M_1M_2}}\left( I^{u}_{1f} -  \frac{I^{u}_{2f} K_1^0}{2\sqrt{M_1 M_2}}\right)
\end{eqnarray}

\begin{eqnarray}\nonumber
V_{u2}&=& \left(I^{u}_{1f} + \frac{I^{u}_{2f} K_1^0}{2\sqrt{M_1 M_2}}\right)\left(I^{u}_{1i} + \frac{I^u_{2i} K_2^0}{2\sqrt{M_1 M_2}}\right) \\\nonumber
&+& \frac{|\vec{K}_1|^2 I^{u}_{2f}}{4\sqrt{M_1M_2}}\left( I^{u}_{1i} +  \frac{I^{u}_{2i} K_2^0}{2\sqrt{M_1 M_2}}\right)\\
&+& \frac{|\vec{K}_2|^2 I^{u}_{2i}}{4\sqrt{M_1M_2}}\left( I^{u}_{1f} +  \frac{I^{u}_{2f} K_1^0}{2\sqrt{M_1 M_2}}\right)\\\nonumber
V_{u3}&=& \left(I^u_{1f} + \frac{I^u_{2f} K_1^0}{2\sqrt{M_1 M_2}}\right)\left(I^u_{1i} - \frac{I^u_{2i}\left( E_1 - M_x\right)}{2\sqrt{M_1 M_2}}\right) \\
&-& \frac{|\vec{K}_1|^2 }{8 M_1M_2}I^u_{2f} I^u_{2i}
\\\nonumber
V_{u4}&=& \left(I^u_{1f} + \frac{I^u_{2f} \left( E_2 + M_x\right)}{2\sqrt{M_1 M_2}}\right)\left(I^u_{1i} - \frac{I^u_{2i}K_2^0}{2\sqrt{M_1 M_2}}\right) \\
 &-& \frac{|\vec{K}_2|^2 }{8 M_1M_2}I^u_{2f} I^u_{2i}
\\\nonumber
V_{u5}&=& \left(I^u_{1f} - \frac{I^u_{2f} K_1^0}{2\sqrt{M_1 M_2}}\right)\left(I^u_{1i} + \frac{I^u_{2i}\left( E_1 + M_x\right)}{2\sqrt{M_1 M_2}}\right) \\
&-& \frac{|\vec{K}_1|^2 }{8 M_1M_2}I^u_{2f} I^u_{2i}\\\nonumber
V_{u6}&=& \left(I^u_{1f} - \frac{I^u_{2f}\left( E_2 - M_x\right) }{2\sqrt{M_1 M_2}}\right)\left(I^u_{1i} + \frac{I^u_{2i} K_2^0}{2\sqrt{M_1 M_2}}\right) \\
&-& \frac{|\vec{K}_2|^2 }{8 M_1M_2}I^u_{2f} I^u_{2i}\\
V_{u7}&=& I^u_{2i}\left(-I^u_{1f} + \frac{I^u_{2f} K_1^0 }{2\sqrt{M_1 M_2}}\right)
\\
V_{u8}&=& I^u_{2f}\left(-I^u_{1i} + \frac{I^u_{2i} K_2^0 }{2\sqrt{M_1 M_2}}\right)
\\
V_{u9}&=& -I^u_{1f}I^u_{2i} - I^u_{2f}\left(I^u_{1i} - \frac{I^u_{2i} (\sqrt{s} -M_x) }{2\sqrt{M_1 M_2}}\right)
\\
V_{u10}&=&  I^u_{2i}\left(I^u_{1f} + \frac{I^u_{2f} K_1^0 }{2\sqrt{M_1 M_2}}\right)
\\
V_{u11}&=&  I^u_{2f}\left(I^u_{1i} + \frac{I^u_{2i} K_2^0 }{2\sqrt{M_1 M_2}}\right)
\\
V_{u12}&=& I^u_{1f}I^u_{2i} + I^u_{2f}\left(I^u_{1i} + \frac{I^u_{2i} (\sqrt{s} + M_x) }{2\sqrt{M_1 M_2}}\right)\label{u12}
\end{eqnarray}

Here, as in the main part of this manuscript, $M_x$ is the mass of the baryon exchanged, $M_1$ ($M_2$), $E_1$ ($E_2$), represent the mass and energy of the incoming (outgoing) baryon in the process  and $K_1^0$ ($K_2^0$), $\vec{K}_1$ ($\vec{K}_2$) represent the energy and three-momentum of the incoming (outgoing) meson. The products of different isospin coefficients; $I^u_{1i}$, $I^u_{1f}$, $I^u_{2i}$, $I^u_{2f}$ are given in Tables~\ref{i1fi1iu}-\ref{i2fi2iu}.  
\begin{table} [h!]
\caption{ The product $I^u_{1f}I^u_{1i}$  appearing in Eqs~(\ref{u1}-\ref{u12}).} \label{i1fi1iu}
\begin{ruledtabular}
\begin{tabular}{cccccc}
&$\rho \Xi$&$\omega \Xi$&$\phi \Xi$&$\bar K^* \Sigma$&$\bar K^* \Lambda$\\
\hline
$\rho \Xi$&$-\frac{1}{4}$&$-\frac{\sqrt{3}}{4}$&$ -\sqrt{\frac{3}{2}}$&$1$&$0$\\
$\omega \Xi$&&$\frac{1}{4}$&$\frac{1}{\sqrt{2}}$&$-\frac{\sqrt{3}}{2}$&$\frac{\sqrt{3}}{2}$ \\
$\phi \Xi$&&& 2 &$-\frac{\sqrt{3}}{2\sqrt{2}}$&$\frac{\sqrt{3}}{2\sqrt{2}}$ \\
$\bar K^* \Sigma$&&&&$-\frac{1}{4}$&$-\frac{3}{4}$\\
$\bar K^*  \Lambda$& & &&&$\frac{3}{4}$\\
\end{tabular}
\end{ruledtabular}
\end{table}
\begin{table} [h!]
\caption{The product $I^u_{1f}I^u_{2i}$  appearing in Eqs~(\ref{u1}-\ref{u12}).} \label{i1fi2iu}
\begin{ruledtabular}
\begin{tabular}{cccccc}
&$\rho \Xi$&$\omega \Xi$&$\phi \Xi$&$\bar K^* \Sigma$&$\bar K^* \Lambda$\\
\hline
$\rho \Xi$&$\frac{(D - F)}{4}$&$\frac{\sqrt{3}(D - F)}{4} $&$ \sqrt{\frac{3}{2}}(D - F)$&$F - \frac{D}{2}$&$\frac{D}{2}$\\
$\omega \Xi$&&$\frac{(F - D)}{4}$&$\frac{(F - D)}{\sqrt{2}}$&$ -\frac{\sqrt{3}F}{2}$&$\sqrt{3}\frac{\left(F - \frac{2D}{3}\right)}{2}$ \\
$\phi \Xi$&&&$2F$&$\frac{1}{2}\sqrt{\frac{3}{2}}(D - F)$&$\frac{(D + 3F)}{2\sqrt{6}}$ \\
$\bar K^* \Sigma$&&&&$\frac{(D - F)}{4}$&$-\frac{(D + 3F)}{4}$\\
$\bar K^*  \Lambda$& & &&&$\frac{(D + 3F)}{4}$\\
\end{tabular}
\end{ruledtabular}
\end{table}
\begin{table} [h!]
\caption{ The product $I^u_{2f}I^u_{1i}$  appearing in Eqs~(\ref{u1}-\ref{u12}).} \label{i2fi1iu}
\begin{ruledtabular}
\begin{tabular}{cccccc}
&$\rho \Xi$&$\omega \Xi$&$\phi \Xi$&$\bar K^* \Sigma$&$\bar K^* \Lambda$\\
\hline
$\rho \Xi$&$\frac{(D - F)}{4}$&$\frac{\sqrt{3}(D - F)}{4} $&$ -\sqrt{\frac{3}{2}} F$&$(D + F)$&$0$\\
$\omega \Xi$&&$\frac{(F - D)}{4}$&$\frac{F}{\sqrt{2}}$&$ -\frac{\sqrt{3}(D + F)}{2}$&$-\frac{(D - 3F)}{2 \sqrt{3}}$ \\
$\phi \Xi$&&&$2F$&$-\frac{1}{2}\sqrt{\frac{3}{2}}(D+F)$&$-\frac{D- 3F}{2\sqrt{6}}$ \\
$\bar K^* \Sigma$&&&&$\frac{(D - F)}{4}$&$\frac{3(D  - F)}{4}$\\
$\bar K^*  \Lambda$& & &&&$\frac{(D + 3F)}{4}$\\
\end{tabular}
\end{ruledtabular}
\end{table}
\begingroup
\squeezetable
 \begin{table} [h!]
\caption{The product $I^u_{2f}I^u_{2i}$  appearing in Eqs~(\ref{u1}-\ref{u12}).} \label{i2fi2iu}
\begin{ruledtabular}
\begin{tabular}{cccccc}
&$\rho \Xi$&$\omega \Xi$&$\phi \Xi$&$\bar K^* \Sigma$&$\bar K^* \Lambda$\\
\hline
$\rho \Xi$&$-\frac{(D - F)^2}{4}$&$-\frac{\sqrt{3}(D - F)^2}{4} $&$ \sqrt{\frac{3}{2}} F(D - F)$&$\frac{(D^2 + 3DF + 6F^2)}{6}$&$\frac{D(D + F)}{2}$\\
$\omega \Xi$&&$\frac{(D - F)^2}{4}$&$\frac{F (F - D)}{\sqrt{2}}$&$ -\frac{\sqrt{3}F (D + F)}{2}$&$\frac{(D - 3F)(2D - 3F)}{6 \sqrt{3}}$ \\
$\phi \Xi$&&&$2F^2$&$\frac{1}{2}\sqrt{\frac{3}{2}}(D^2 - F^2)$&$-\frac{(D^2 - 9F^2)}{6\sqrt{6}}$ \\
$\bar K^* \Sigma$&&&&$-\frac{(D - F)^2}{4}$&$\frac{(D^2 + 2DF - 3F^2)}{4}$\\
$\bar K^*  \Lambda$& & &&&$\frac{(D + 3F)^2}{12}$\\
\end{tabular}
\end{ruledtabular}
\end{table}
\endgroup

\section{  CONVENTIONS FOLLOWED} \label{conv}

We choose the three-momentum of the incoming meson to be along the positive z-axis, such that
three-momenta of the incoming and the outgoing mesons can be written as
 \begin{eqnarray}\nonumber
\vec{K}_1&=&(0,0,|\vec{K}_1|),\\\nonumber
\vec{K}_2&=&(|\vec{K}_2| {\rm sin~\theta}, 0, |\vec{K}_2| {\rm cos~\theta}).
\end{eqnarray}
In the center of mass system then, we can write the three-momenta of the incoming and the outgoing baryons as: 
\begin{eqnarray}\nonumber
\vec{P}_1&=&(0,0,-|\vec{K}_1|),\\\nonumber
\vec{P}_2&=&(-|\vec{K}_2| {\rm sin~\theta}, 0,- |\vec{K}_2| {\rm cos~\theta}).
\end{eqnarray}
The corresponding four vectors are written as:\\ $K_1 = (K_1^0, \vec{K}_1)$, $K_2 = (K_2^0, \vec{K}_1)$, $P_1 = (E_1, -\vec{K}_1)$, $P_2 = (E_2, -\vec{K}_2)$.\\

\noindent
The three-polarization vector of the incoming meson is taken as
\begin{eqnarray}
\vec{\epsilon}_{1,  \pm} &=& \mp \frac{1}{\sqrt{2}} \left(\vec{e}_x \pm \vec{e}_y \right),\\\nonumber
 \vec{\epsilon}_{1, 0}  &=& \vec{e}_z,
\end{eqnarray}
and the corresponding four vectors are 
\begin{eqnarray}
\epsilon_{1, \pm} &=& \left( 0, \vec{\epsilon}_{1, \pm} \right),\\\nonumber
\epsilon_{1, 0} &=& \left( \dfrac{\mid \vec{K}_1\mid}{m_1}, \vec{\epsilon}_{1, 0} \dfrac{K_1^0}{m_1}\right).
\end{eqnarray} 
In this way, we have the Lorenz gauge condition $K_1~\cdot~\epsilon_1~(\lambda)~=~0$, as well as  the following  conditions,  satisfied: 

\begin{align}\label{cond1}\vec{\epsilon}_{1, \lambda} \cdot \vec{\epsilon}_{1, \lambda^\prime} = \delta_{\lambda, \lambda^\prime}\end{align}
\begin{align}\label{cond2}\sum\limits_{\lambda = 1}^{3} \epsilon_{1\mu, \lambda}^* \epsilon_{1\nu, \lambda} = - g_{\mu \nu} + \dfrac{K_{1\mu}K_{1\nu}}{m_1^2}\end{align}
\begin{align}\label{cond3}\sum\limits_{\lambda = 0}^{3} g_{\lambda \lambda} \epsilon_{1\mu, \lambda}^* \epsilon_{1\nu, \lambda} =  g_{\mu \nu} \end{align}

The three-polarization vector of the outgoing meson is obtained by rotating the $\vec{\epsilon}_{1, \lambda}$, such that $\vec{\epsilon}_{2, 0}$ is parallel to $\vec{K}_2$, which gives \cite{pennerthesis}
\small{
\begin{eqnarray}\nonumber
\vec{\epsilon}_{2, +} = \frac{1}{\sqrt{2}} \left(\begin{array}{c} - {\rm cos~\theta} \\ - i \\  {\rm sin~\theta} \end{array}\right); \vec{\epsilon}_{2, 0} = \left(\begin{array}{c}  {\rm sin~\theta} \\ 0 \\  {\rm cos~\theta} \end{array}\right);  \vec{\epsilon}_{2, -} = \frac{1}{\sqrt{2}} \left(\begin{array}{c}  {\rm cos~\theta} \\ -i  \\  -{\rm sin~\theta} \end{array}\right)
\end{eqnarray}}
and the corresponding four vectors are:
\begin{eqnarray}
\epsilon_{2, \pm} &=& \left(0, \vec{\epsilon}_{2, \pm}\right) ;\\\nonumber
\epsilon_{2, 0} &=& \left(\frac{|\vec{K}_2|}{m_2}, \vec{\epsilon}_{2, 0} \frac{K_2^0}{m_2},\right).
\end{eqnarray}
These four vectors satisfy the Lorenz gauge condition as well as the normalization conditions listed by Eqs~(\ref{cond1}-\ref{cond3}).


\begin{thebibliography}{99}
\bibitem{pdg} 
 J. Beringer  {\it et al.}  (Particle Data Group), Phys.\ Rev.\ D {\bf 86}, 010001 (2012). 
 
   \bibitem{clasx}
  L.~Guo {\it et al.},
  %``Cascade production in the reactions gamma p ---> K+ K+ (X) and gamma p ---> K+ K+ pi- (X),''
  Phys.\ Rev.\ C {\bf 76}, 025208 (2007);
  R.~Schumacher [CLAS Collaboration],
  %``Strangeness physics with CLAS at Jefferson Lab,''
  AIP Conf.\ Proc.\  {\bf 1257}, 100 (2010).
  %%CITATION = APCPC,1257,100;%%
 

  
\bibitem{jparc}
  T.~Nagae,
  %``The J-PARC project,''
  Nucl.\ Phys.\ A {\bf 805}, 486 (2008).

  
\bibitem{panda}
  M.~F.~M.~Lutz {\it et al.}  [PANDA Collaboration],
  %``Physics Performance Report for PANDA: Strong Interaction Studies with Antiprotons,''
  arXiv:0903.3905 [hep-ex].
   \bibitem{belle}
  K.~Abe {\it et al.}  [Belle Collaboration],
  %``Observation of Cabibbo suppressed and W exchange Lambda+(c) baryon decays,''
  Phys.\ Lett.\ B {\bf 524}, 33 (2002).
  

  \bibitem{babar}
  B.~Aubert {\it et al.} [BaBar Collaboration],
  %``Measurement of the Spin of the Xi(1530) Resonance,''
  Phys.\ Rev.\ D {\bf 78}, 034008 (2008)
 % doi:10.1103/PhysRevD.78.034008
  [arXiv:0803.1863 [hep-ex]]; B.~Aubert {\it et al.}  [BaBar Collaboration],
  %``Measurement of the Mass and Width and Study of the Spin of the $\Xi(1690) $ 0 Resonance from $\Lambda^+_{c} \to \Lambda \bar{K}^0 K^{+}$ Decay at Babar,''
  hep-ex/0607043;  V.~Ziegler,
  %``Hyperon and Hyperon Resonance Properties from Charm Baryon Decays at BaBar,''
  SLAC-R-868.
  
  \bibitem{Dionisi:1978tg} 
  C.~Dionisi {\it et al.} [Amsterdam-CERN-Nijmegen-Oxford Collaboration],
  %``An Enhancement at the $\Sigma \bar{K}$ Threshold 1680-{MeV} Observed in $K^- p$ Reactions at 4.2-{GeV}/$c$,''
  Phys.\ Lett.\ B {\bf 80}, 145 (1978).
  %doi:10.1016/0370-2693(78)90329-5
  
  \bibitem{Pervin:2007wa} 
  M.~Pervin and W.~Roberts,
  %``Strangeness -2 and -3 baryons in a constituent quark model,''
  Phys.\ Rev.\ C {\bf 77}, 025202 (2008)
  %doi:10.1103/PhysRevC.77.025202
  [arXiv:0709.4000 [nucl-th]].

%\cite{Xiao:2013xi}
\bibitem{Xiao:2013xi} 
  L.~Y.~Xiao and X.~H.~Zhong,
  %``$\Xi$ baryon strong decays in a chiral quark model,''
  Phys.\ Rev.\ D {\bf 87}, 094002 (2013)
  %doi:10.1103/PhysRevD.87.094002
  [arXiv:1302.0079 [hep-ph]].

  
    %\cite{Sekihara:2015qqa}
\bibitem{Sekihara} 
  T.~Sekihara,
  %``$\Xi (1690)$ as a $\bar{K} \Sigma$ molecular state,''
  PTEP {\bf 2015}, 091D01 (2015)
  %doi:10.1093/ptep/ptv129
  [arXiv:1505.02849 [hep-ph]].
  %%CITATION = doi:10.1093/ptep/ptv129;%%
  %3 citations counted in INSPIRE as of 04 May 2016
  
    %\cite{Tornqvist:1991ks}
\bibitem{Tornqvist:1991ks} 
  N.~A.~Tornqvist,
  %``Possible large deuteron - like meson meson states bound by pions,''
  Phys.\ Rev.\ Lett.\  {\bf 67}, 556 (1991).
  %doi:10.1103/PhysRevLett.67.556
  %%CITATION = doi:10.1103/PhysRevLett.67.556;%%
  
    %\cite{Ramos:2002xh}
\bibitem{ramosXi} 
  A.~Ramos, E.~Oset and C.~Bennhold,
  %``On the spin, parity and nature of the Xi(1620) resonance,''
  Phys.\ Rev.\ Lett.\  {\bf 89}, 252001 (2002)
%  doi:10.1103/PhysRevLett.89.252001
  [nucl-th/0204044].
  %%CITATION = doi:10.1103/PhysRevLett.89.252001;%%
  %70 citations counted in INSPIRE as of 04 May 2016
    
 \bibitem{Kolomeitsev:2003kt} 
  E.~E.~Kolomeitsev and M.~F.~M.~Lutz,
  %``On baryon resonances and chiral symmetry,''
  Phys.\ Lett.\ B {\bf 585}, 243 (2004)
  %doi:10.1016/j.physletb.2004.01.066
  [nucl-th/0305101].
  
  \bibitem{GarciaRecio:2003ks} 
  C.~Garcia-Recio, M.~F.~M.~Lutz and J.~Nieves,
  %``Quark mass dependence of s wave baryon resonances,''
  Phys.\ Lett.\ B {\bf 582}, 49 (2004)
 % doi:10.1016/j.physletb.2003.11.073
  [nucl-th/0305100].
  
    \bibitem{ramosvb}
E.~Oset and A.~Ramos,
  %``Dynamically generated resonances from the vector octet-baryon octet interaction,''
 Eur.\ Phys.\ J.\ A {\bf 44} (2010) 445.
  
  %\cite{Gamermann:2011mq}
\bibitem{Gamermann} 
  D.~Gamermann, C.~Garcia-Recio, J.~Nieves and L.~L.~Salcedo,
  %``Odd Parity Light Baryon Resonances,''
  Phys.\ Rev.\ D {\bf 84}, 056017 (2011)
 % doi:10.1103/PhysRevD.84.056017
  [arXiv:1104.2737 [hep-ph]].
  %%CITATION = doi:10.1103/PhysRevD.84.056017;%%
  %37 citations counted in INSPIRE as of 04 May 2016
  
  %\cite{PavonValderrama:2011gp}
\bibitem{Pavon} 
  M.~Pavon Valderrama, J.~J.~Xie and J.~Nieves,
  %``Are there three Xi(1950) states?,''
  Phys.\ Rev.\ D {\bf 85}, 017502 (2012)
 % doi:10.1103/PhysRevD.85.017502
  [arXiv:1111.2218 [hep-ph]].
  %%CITATION = doi:10.1103/PhysRevD.85.017502;%%
  %4 citations counted in INSPIRE as of 04 May 2016

%\cite{Nakayama:2012zp}
\bibitem{Nakayama} 
  K.~Nakayama, Y.~Oh and H.~Haberzettl,
  %``Model-independent determination of the parity of $\Xi$ hyperons,''
  Phys.\ Rev.\ C {\bf 85}, 042201 (2012)
  %doi:10.1103/PhysRevC.85.042201
  [arXiv:1201.5598 [hep-ph]].
  %%CITATION = doi:10.1103/PhysRevC.85.042201;%%
  
    \bibitem{javi}
  E.~J.~Garzon and E.~Oset,
  %``Effects of pseudoscalar-baryon channels in the dynamically generated vector-baryon resonances,''
  Eur.\ Phys.\ J.\ A {\bf 48} (2012) 5.
  

  %\cite{Oh:2011np}
\bibitem{Oh} 
  Y.~Oh,
  %``Hyperons analogous to the \Lambda(1405),''
  Few Body Syst.\  {\bf 54}, 411 (2013)
  %doi:10.1007/s00601-012-0401-7
  [arXiv:1111.7044 [nucl-th]].
  %%CITATION = doi:10.1007/s00601-012-0401-7;%%
  %1 citations counted in INSPIRE as of 04 May 2016

  %\cite{Gay:1976sr}
\bibitem{Gay:1976sr} 
  J.~B.~Gay {\it et al.} [Amsterdam-CERN-Nijmegen-Oxford Collaboration],
  %``Production and Decay of xi* (1820) in K- p Reactions at 4.2-GeV/c,''
  Phys.\ Lett.\ B {\bf 62}, 477 (1976).
  %doi:10.1016/0370-2693(76)90688-2
  
  %\cite{Vorobev:1979as}
\bibitem{Vorobev:1979as} 
  A.~P.~Vorobev {\it et al.} [French-Soviet and CERN-Soviet Collaborations],
  %``A STUDY OF anti-LAMBDA pi+, anti-LAMBDA K+ AND anti-LAMBDA p PRODUCTION IN 32-GeV/c K+ p INTERACTIONS,''
  Nucl.\ Phys.\ B {\bf 158}, 253 (1979).
 % doi:10.1016/0550-3213(79)90164-0
  %%CITATION = doi:10.1016/0550-3213(79)90164-0;%%
  
  \bibitem{Hemingway:1977uw} 
  R.~J.~Hemingway {\it et al.} [Amsterdam-CERN-Nijmegen-Oxford Collaboration],
  %``Xi* (2030) Production in K- p Reactions at 4.2-GeV/c,''
  Phys.\ Lett.\ B {\bf 68}, 197 (1977).
  %doi:10.1016/0370-2693(77)90200-3
  %%CITATION = doi:10.1016/0370-2693(76)90688-2;%%
  %%CITATION = %doi:10.1103/PhysRevD.87.094002;%%


\bibitem{hades}
G.~Agakishiev {\it et al.}  [HADES Collaboration],
  %``Deep sub-threshold Xi- production in Ar+KCl reactions at 1.76A-GeV,''
  Phys.\ Rev.\ Lett.\  {\bf 103}, 132301 (2009);
    G.~Agakishiev {\it et al.} [HADES Collaboration],
  %``Statistical model analysis of hadron yields in proton-nucleus and heavy-ion collisions at SIS 18 energiesStatistical hadronization model analysis of hadron yields in p + Nb and Ar + KCl at SIS18 energies,''
  Eur.\ Phys.\ J.\ A {\bf 52}, no. 6, 178 (2016);
 % doi:10.1140/epja/i2016-16178-x
 % [arXiv:1512.07070 [nucl-ex]].
  %%CITATION = doi:10.1140/epja/i2016-16178-x;%%
  G.~Agakishiev {\it et al.} [HADES Collaboration],
  %``Statistical model analysis of hadron yields in proton-nucleus and heavy-ion collisions at SIS 18 energiesStatistical hadronization model analysis of hadron yields in p + Nb and Ar + KCl at SIS18 energies,''
  Eur.\ Phys.\ J.\ A {\bf 52}, no. 6, 178 (2016).
%  doi:10.1140/epja/i2016-16178-x
 % [arXiv:1512.07070 [nucl-ex]].
 

  
  \bibitem{vbvb}
  K.~P.~Khemchandani, H.~Kaneko, H.~Nagahiro and A.~Hosaka,
  %``Vector meson-Baryon dynamics and generation of resonances,''
  Phys.\ Rev.\ D {\bf 83} (2011)  114041.
   
   
        \bibitem{jidohosaka}
%\cite{Jido:2002yz}
%\bibitem{Jido:2002yz}
  D.~Jido, A.~Hosaka, J.~C.~Nacher, E.~Oset, A.~Ramos,
  %``Magnetic moments of the Lambda(1405) and Lambda(1670) resonances,''
  Phys.\ Rev.\  {\bf C66}, 025203 (2002).
  
  \bibitem{Penner:2002ma} 
  G.~Penner and U.~Mosel,
  %``Vector meson production and nucleon resonance analysis in a coupled channel approach for energies m(N) less than s**(1/2) less than 2-GeV. 1. Pion induced results and hadronic parameters,''
  Phys.\ Rev.\ C {\bf 66}, 055211 (2002).
 
  \bibitem{Penner:2002md} 
  G.~Penner and U.~Mosel,
  %``Vector meson production and nucleon resonance analysis in a coupled channel approach for energies m(N) less than S**(1/2) less than 2-GeV. 2. Photon induced results,''
  Phys.\ Rev.\ C {\bf 66}, 055212 (2002)
  doi:10.1103/PhysRevC.66.055212


   \bibitem{Maris:1999nt}
  P.~Maris and P.~C.~Tandy,
  %``Bethe-Salpeter study of vector meson masses and decay constants,''
  Phys.\ Rev.\ C {\bf 60} (1999) 055214.
  %%CITATION = NUCL-TH/9905056;%%
  
  \bibitem{Maris:1999sz}
  P.~Maris,
  %``Dyson-Schwinger studies of meson masses and decay constants,''
  Nucl.\ Phys.\ A {\bf 663} (2000) 621.
  
  \bibitem{raquelgeng} 
  L.~S.~Geng, R.~Molina and E.~Oset,
  %``On the chiral covariant approach to $\rho\rho$ scattering,''
  arXiv:1612.07871 [nucl-th].
 
        \bibitem{pbvb}
  K.~P.~Khemchandani, A.~Martinez Torres, H.~Kaneko, H.~Nagahiro and A.~Hosaka,
  %``Coupling vector and pseudoscalar mesons to study baryon resonances,''
  Phys.\ Rev.\ D {\bf 84} (2011) 094018.
    \bibitem{hyperons}
    K.~P.~Khemchandani, A.~Martinez Torres, H.~Nagahiro and A.~Hosaka,
  %``Negative parity $\Lambda$ and $\Sigma$ resonances coupled to pseudoscalar and vector mesons,''
  Phys.\ Rev.\ D {\bf 85} (2012) 114020.
      \bibitem{more}
   K.~P.~Khemchandani, A.~Martinez Torres, H.~Nagahiro and A.~Hosaka,
  %``Role of vector and pseudoscalar mesons in understanding $1/2^-$ $N^*$ and $\Delta$ resonances,''
  Phys.\ Rev.\ D {\bf 88} (2013) 114016; K.~P.~Khemchandani, A.~Martinez Torres, H.~Nagahiro and A.~Hosaka,
  %``$N^*$'s and $\Delta$'s generated in vector, pseudoscalar meson-baryon systems,''
  Int.\ J.\ Mod.\ Phys.\ Conf.\ Ser.\  {\bf 26} (2014) 1460060.
  

  
    \bibitem{ecker}
 G.~Ecker, Prog.\ Part.\ Nucl.\ Phys. {\bf 35} (1995) 1.
 
 \bibitem{pich}
 A.~Pich, Rep.\ Prog.\ Phys. {\bf 58} (1995) 563.
 

  
    \bibitem{osetramos}
  E.~Oset and A.~Ramos,
  %``Nonperturbative chiral approach to s wave anti-K N interactions,''
  Nucl.\ Phys.\ A {\bf 635} (1998) 99.
  
     \bibitem{oller} 
  J.~A.~Oller and E.~Oset,
  %``Chiral symmetry amplitudes in the S wave isoscalar and isovector channels and the sigma, f0(980), a0(980) scalar mesons,''
  Nucl.\ Phys.\ A {\bf 620} (1997) 438 
  [Erratum-ibid.\ A {\bf 652} (1999) 407].
  


  
   \bibitem{hyodo}   
      T.~Hyodo, D.~Jido and A.~Hosaka,
  %``Compositeness of dynamically generated states in a chiral unitary approach,''
  Phys.\ Rev.\ C {\bf 85}, 015201 (2012).
  
  \bibitem{Oller:2000fj} 
  J.~A.~Oller and U.~G.~Meissner,
  %``Chiral dynamics in the presence of bound states: Kaon nucleon interactions revisited,''
  Phys.\ Lett.\ B {\bf 500}, 263 (2001).
  
        \bibitem{inoue}
    T.~Inoue, E.~Oset and M.~J.~Vicente Vacas,
  Phys.\ Rev.\  C {\bf 65}, 035204 (2002).
  
  \bibitem{MartinezTorres:2012du} 
  A.~Martinez Torres, K.~P.~Khemchandani, F.~S.~Navarra, M.~Nielsen and E.~Oset,
  %``The Role of $f_0(1710)$ in the $\phi \omega$ Threshold Peak of $J/\Psi \to \gamma \phi \omega$,''
  Phys.\ Lett.\ B {\bf 719}, 388 (2013);   A.~Martinez Torres, K.~P.~Khemchandani, F.~S.~Navarra, M.~Nielsen and E.~Oset,
  %``Reanalysis of the $e^+ e^- \to (D^* \bar D^*)^{\pm} \pi^{\mp}$ reaction and the claim for the $Z_c (4025)$ resonance,''
  Phys.\ Rev.\ D {\bf 89}, no. 1, 014025 (2014).
 % doi:10.1103/PhysRevD.89.014025
 % doi:10.1016/j.physletb.2013.01.036
  
   \bibitem{selfenergy} 
  B.~Julia-Diaz, T.-S.~H.~Lee, A.~Matsuyama and T.~Sato,
  %``Dynamical coupled-channel model of pi N scattering in the W <= 2-GeV nucleon resonance region,''
  Phys.\ Rev.\ C {\bf 76}, 065201 (2007); H.~Kamano, B.~Julia-Diaz, T.-S.~H.~Lee, A.~Matsuyama and T.~Sato,
  %``Dynamical coupled-channels study of pi N ---> pi pi N reactions,''
  Phys.\ Rev.\ C {\bf 79}, 025206 (2009);   N.~Suzuki, T.~Sato and T.-S.~H.~Lee,
  %``Extraction of Resonances from Meson-Nucleon Reactions,''
  Phys.\ Rev.\ C {\bf 79}, 025205 (2009); 
  N.~Suzuki, T.~Sato and T.-S.~H.~Lee,
  %``Extraction of Electromagnetic Transition Form Factors for Nucleon Resonances within a Dynamical Coupled-Channels Model,''
  Phys.\ Rev.\ C {\bf 82}, 045206 (2010).
  
  \bibitem{Borasoy}
    B.~Borasoy, R.~Nissler and W.~Weise,
  %``Chiral dynamics of kaon-nucleon interactions, revisited,''
  Eur.\ Phys.\ J.\ A {\bf 25}, 79 (2005).
  %doi:10.1140/epja/i2005-10079-1
  
  \bibitem{Alt:2003vb} 
  C.~Alt {\it et al.} [NA49 Collaboration],
  %``Observation of an exotic S = -2, Q = -2 baryon resonance in proton proton collisions at the CERN SPS,''
  Phys.\ Rev.\ Lett.\  {\bf 92}, 042003 (2004)
  %doi:10.1103/PhysRevLett.92.042003
  
  \bibitem{Abelev:2014qqa} 
  B.~B.~Abelev {\it et al.} [ALICE Collaboration],
  %``Production of $\Sigma(1385)^{\pm}$ and $\Xi(1530)^{0}$ in proton-proton collisions at $\sqrt{s}=$ 7 TeV,''
  Eur.\ Phys.\ J.\ C {\bf 75}, 1 (2015)
  %doi:10.1140/epjc/s10052-014-3191-x
  
    \bibitem{hics}
  E.~E.~Kolomeitsev, B.~Tomasik and D.~N.~Voskresensky,
  %``Strangeness Balance in HADES Experiments and the $\Xi^-$ Enhancement,''
  Phys.\ Rev.\ C {\bf 86}, 054909 (2012).
  
  \bibitem{Chen:2003nm} 
  L.~W.~Chen, C.~M.~Ko and Y.~Tzeng,
  %``Cascade production in heavy ion collisions at SIS energies,''
  Phys.\ Lett.\ B {\bf 584}, 269 (2004).
 % doi:10.1016/j.physletb.2004.01.055
  %[nucl-th/0312009].
  %%CITATION = doi:10.1016/j.physletb.2004.01.055;%%
  %13 citations counted in INSPIRE as of 05 Jul 2016

  
  \bibitem{Li:2012bga} 
  F.~Li, L.~W.~Chen, C.~M.~Ko and S.~H.~Lee,
  %``Contributions of hyperon-hyperon scattering to subthreshold cascade production in heavy ion collisions,''
  Phys.\ Rev.\ C {\bf 85}, 064902 (2012).
 % doi:10.1103/PhysRevC.85.064902
 % [arXiv:1204.1327 [nucl-th]].
 
 

 
 \bibitem{pennerthesis}
 G.~Penner, Ph. D thesis, University of Giessen, 2002.
\end{thebibliography}
\end{document}